\documentclass{aa}  

\newcommand{\gsim}{\ \raise-2.truept\hbox{\rlap{\hbox{$\sim$}}\raise 5.truept\hbox{$>$}\ }}

\newcommand{\leda}{PGC\,087327}
\newcommand{\ksm}{km~s$^{-1}$~Mpc$^{-1}$}
\newcommand{\g}{$g_{F475W}\ $}
\newcommand{\V}{$V_{F606W}\ $}

\usepackage{graphicx}
\usepackage{lscape}
\usepackage{placeins}
\usepackage{tablefootnote}

\usepackage{txfonts}
\begin{document}

   \title{A study of globular clusters in a lenticular galaxy in Hydra\,I from deep HST/ACS photometry}

   \subtitle{}

   \author{Nandini Hazra \inst{1,2,3}
    \and 
    Michele Cantiello \inst{3}
    \and 
    Gabriella Raimondo \inst{3}
    \and
    Marco Mirabile \inst{3,4}
    \and
    John\,P.\,Blakeslee \inst{6}
    \and
    Marica Branchesi \inst{1,2,3}
    \and
    Enzo Brocato \inst{3,5}}
          

   \institute{Gran Sasso Science Institute, L'Aquila, Italy
        \and
            INFN Laboratori Nazionali del Gran Sasso, L’Aquila, Italy
         \and
             INAF - Osservatorio Astronomico d’Abruzzo, Teramo, Italy
             \and
             University of Naples Federico II, Naples, Italy
             \and
             INAF - Osservatorio Astronomico di Roma,  Monte Porzio Catone (Roma), Italy
             \and
             NSF’s NOIRLab, Tucson, AZ 85719, USA
             }


  \abstract
   {}
      {We take advantage of exquisitely deep optical imaging data from HST/ACS in the F475W (\g) and F606W (\V) bands, to study the properties of the globular cluster (GC) population in the intermediate mass lenticular galaxy \leda, in the Hydra\,I galaxy cluster.}
      {We inspect the photometric (magnitudes and color) and morphometric (compactness, elongation, etc.) properties of sources lying in an area of $\sim19\times19$ kpc centered on \leda \, and compare them with four neighbouring fields over the same HST/ACS mosaic. This allowed us to identify a list of GC candidates and to inspect their properties using a background decontamination method.}
        {Relative to four comparison fields, \leda \, shows a robust overdensity of GCs, $N_{GC}=82\pm9$. At the estimated magnitude of the galaxy, this number implies a specific frequency of $S_N=1.8\pm0.7$. In spite of the short wavelength interval available with the \g and \V passbands, the color distribution  shows a clear bimodality with a blue peak at $\langle g_{F475W}{-}V_{F606W} \rangle =0.47\pm0.05$ mag and a red peak at $\langle g_{F475W}{-}V_{F606W}\rangle =0.62\pm0.03$ mag. We also observe the typical steeper slope of the radial distribution of red GCs relative to blue ones. Thanks to the unique depth of the available data, we characterize the GC luminosity function (GCLF) well beyond the expected GCLF turn-over. We find $g^{TOM}_{F475W} = 26.54\pm0.10$ mag and $V^{TOM}_{F606W} = 26.08 \pm 0.09$ mag, which after calibration yields a distance of $D_{GCLF} = 56.7 \pm 4.3(statistical) \pm 5.2(systematic)$ Mpc.}
   {}


   \maketitle
%

\section{Introduction}
\label{sec:intro}
%


Globular clusters (GC) are dense stellar systems, with typically old ages, found ubiquitously in massive galaxies and spanning a wide range of magnitudes \citep[e.g. ][]{brodie06}. The small number of available spectroscopic studies of extragalactic GCs inferred that the majority of them have ages  comparable to galactic GCs \citep[e.g., ][]{cohen98,beasley00,kuntschner02}, namely $t\geq 10$ Gyr \citep[][]{carretta00}. In all cases where the population of GCs in a spheroidal galaxy, either elliptical or lenticular, was observed with an intermediate age component ($t\sim3-6$ Gyr), this only composed a small fraction of the GC population, mostly in merger remnants like NGC\,3610 or NGC\,1316 \citep[][]{goudfrooij01b,brodie06,bassino17}.

Throughout this paper we will focus only on the old GCs component. They are often some of the most luminous non-transient objects in a galaxy and exhibit a variety of properties (magnitudes, colors, radial distributions, sizes and so on) that are used as tracers of galaxy formation and evolution \citep[e.g.][]{brodie06}. These old GC systems have been used as distance indicators since \citet[][]{hanes77} due to the fact that they have a nearly universal Gaussian luminosity function, with a peak at a constant absolute magnitude of $M_V^{TO}\sim-7.6$ mag \citep[][]{iskrenpuzia09}, known as the turn-over magnitude. The near-universality of the globular cluster luminosity function (GCLF) in optical and near-IR bands, has prompted its use as a standard candle to act as a secondary distance indicator \citep[][]{ferrarese00a}.

Old GC populations typically exhibit a bimodal color distribution, with a blue (metal-poor) and a red (metal-rich) peak. This has historically been attributed to hierarchical formation giving rise to two distinct GC sub-populations with different peak metallicities \citep[][]{brodie06}. Additionally, while the red GC system generally shows a radial profile which roughly matches with the galaxy field star profile \citep{harris09}, the blue, more metal-poor GCs, are often observed to be less concentrated close to the galaxy core and  more numerous than the red GCs at larger radii. This seems to be indicative of an {\it in-situ} red GCs population, and of a blue population acquired through galaxy mergers and tidal events \citep[][]{forbes11}.

However, recent works  \citep[][among others]{yoon06, richtler06, cantiello07c, cantiello07d, cantiello12b} have shown that the bimodality could also arise from a continuous metallicity distribution with a radial gradient, combined with non-linear color-metallicity relations. This could point to stochastic processes in galaxy formation, without requiring two major events or mechanisms to generate the two observed sub-populations.

In this work, we take advantage of the exquisite resolution of HST/ACS data, combined with the extremely deep observations of NGC\,3314A/B, to characterize the GC system in \leda. In particular, we analyze the color and radial distributions, and the luminosity function of the old GCs in this galaxy.

\leda~is an intermediate-mass galaxy (see Sect. \ref{sec:data}) classified as E3/S0, close in projection to NGC\,3314A/B, in the Hydra\,I cluster. Using the flow-corrected peculiar velocity from Cosmicflows-3 \citep{tully16} reported in Table \ref{tab:properties}, and an $H_0\sim73$ \ksm \citep[e.g.][]{blake21, khetan20}, we obtain a preliminary estimate of the distance, $D\sim61$ Mpc. Later in this work, we will use the photometry of GCs to derive a more refined estimate of this distance.

\begin{table}
\caption{Main properties of \leda}             
\label{tab:properties}  
\setlength{\tabcolsep}{3pt}
\centering                          
\begin{tabular}{l l l }        
\hline \\ [-1.5ex]                
Quantity & Value & Reference\\ \\ [-1.5ex]   
\hline  \\ [-1.5ex]  
   $v_{Helio.}$&4114 $km\ s^{-1}$ &\citet[][]{smith00} \\
   $v_{Corr.}$&4449 $km\ s^{-1}$ &\citet[][]{tully16} \\
   Mass &$\sim10^{10}M_{\odot}$ &This work\\
   Dist.($H_0$) & 61 Mpc & Hubble-Lemaitre law \& $v_{Corr.}$ \\ 
   $E_{B-V}$&0.063 &IRSA\\
   $m_B$ & $15.6$ mag$^{a}$& \citet[][]{bernardi02}\\
   $m_{Ks}$ & $11.7$ mag$^{a}$& 2MASS\\
   $m_H$ & $13.3$ mag& 2MASS\\
   $m_z$ & $13.5$ mag  &  $m_{H}$ \& $\langle z{-}H\rangle\sim0.2$ mag \\
   $m_{F606W}$& 14.7 mag & This work\\
   $m_{F475W}$& 15.3 mag & This work\\
    \\ [-1.5ex]
\hline                                   
\end{tabular}\\
\footnotesize{Total magnitudes, $m_{band}$, have not been corrected for extinction.
Magnitudes in the $K_s$, $H$, $z$ bands as well as those estimated in this work (see Sect. \ref{sec:model}) are from fit extrapolation estimates.
\footnotesize{\\Note: $a)$ These magnitudes are in Vegamag, for sake of comparison with the literature used throughout this work.}}
\end{table}

This paper is organised as follows: we describe the observational dataset and target in Section \ref{sec:data}, and the procedures to identify GC candidates in Sect. \ref{sec:gc_sel}. The analysis of the main properties of the GC sample, along with the calibrations and results are outlined in Sect. \ref{sec:analysis}, and the conclusions are summarized in Sect. \ref{sec:conclusions}.

\section{Observational Data}
\label{sec:data}

\subsection{Target properties}

\begin{figure*}
   \centering
   \includegraphics[width=18cm]{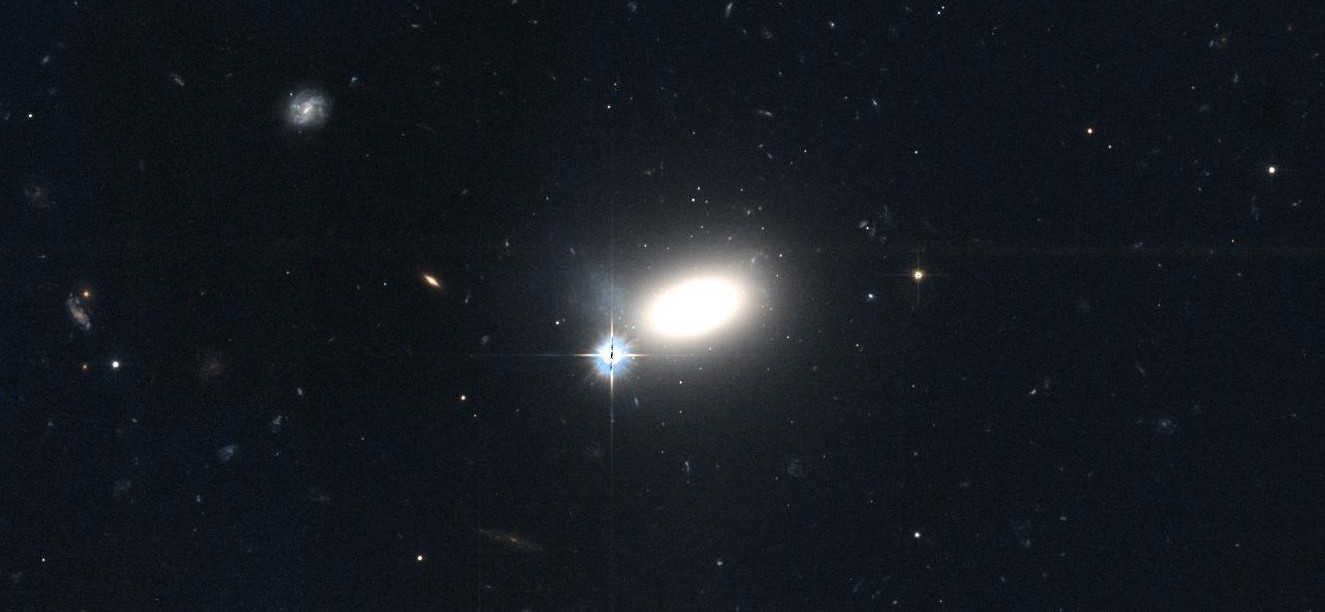}
   \caption{Color composite $2\arcmin \times 1\arcmin$ HST image of \leda. North is up and east is left}.
              \label{fig:closeup}%
\end{figure*}

The main properties of \leda \, are given in Table \ref{tab:properties}.
Figure \ref{fig:closeup} shows the color composite image of the galaxy based on the HST data used in this work.

We derived an approximate estimate of the galaxy mass using stellar population models and empirical relations. For the estimate from models we adopted the colour-mass-luminosity relations of \citet[][see their Tab. 3]{into13} together with the $K_s$ magnitude from 2MASS (the magnitude from fit extrapolation, see Tab. \ref{tab:properties}), the distance from the Hubble-Lemaitre law and a range of $V{-}I$ color of 1-1.25 mag \citep[e.g.][]{tonry01}. With such assumptions we evaluate a total mass in the range $9.85 \leq \log (M/M_{\odot}) \leq 10.2$, depending on the color used. Using the empirical mass-luminosity relation from \citet[][their eq. 2]{cappellari13}, combined with the $K_s$ magnitude and distance, we obtained $\log (M/M_{\odot})\sim 10$. From these estimates, \leda \, appears to be an intermediate mass lenticular galaxy.

For the GCLF calibrations (see Sect. \ref{sec:results}) we need the $B$- and $z$-band magnitudes of the galaxy. We adopted the $m_B$ estimate from \citet[][\leda~is D\,135 in their catalog]{bernardi02}, reported in Table \ref{tab:properties}. For the $z$-band magnitude, we used the 2MASS $m_{H}=13.33$ mag\footnote{Throughout this work, we will always consider the ABmag photometric system, unless specified otherwise.} and a $z{-}H\sim0.2$ color --appropriate for elliptical galaxies, \citet[][]{leechary20}--, thus obtaining $m_{z}=13.53$ mag.

Additionally, we find a $B$-band mass-luminosty ratio of $2.5\leq (M/L)\leq3$ for this galaxy, which compared to the predictions of \citet[][see their Fig. 4]{into13} indicates stellar population ages older than $\sim10$ Gyr for all except the highest metallicities. A field stellar component with metallicities $[Fe/H]\geq0$ is ruled out  by the measured galaxy color (computed later in Sect. \ref{sec:model}, total magnitudes without extinction correction are in Table \ref{tab:properties}), which is observed to be slightly bluer (by $\sim 0.1$ mag) than model predictions for intermediate-age metal rich populations.

\subsection{HST Data} 
The data for this analysis was retrieved from the Hubble Legacy Archive\footnote{See {\url{https://hla.stsci.edu/}}}, and are part of the  gravitational microlensing survey in the NGC\,3314A/B galaxy pair (HST Proposal 9977, PI. D. Bennett). 

The observations were carried out in the $F475W$ and $F606W$ passbands (hereafter also referenced as \g and \V, respectively). We downloaded the combined images based on the standard HST drizzling and calibration pipeline.

Table \ref{tab:observations} provides information about the observed dataset and the ABmag zeropoints we adopted. The full frame is $\sim5\arcmin$ wide, and centered on the double spiral galaxy NGC\,3314A/B. The original ACS resolution is ~0.05$\arcsec/pix$ but the mosaic downloaded from the Hubble Legacy Archive has been drizzled to a lower pixel resolution of $0.04\arcsec/pix$ owing to the very large number of dithered exposures ($N_{exp}=120$). In order to isolate the GCs around our target, we chose a cutout of the frame centered on \leda, having an approximate width of $1.1\arcmin\times1.1\arcmin$ ($1600\times1600$ pixels $\approx 19\times 19$ kpc at the assumed galaxy distance of 61 Mpc). The galaxy lies at $\sim1.7\arcmin$ from the overlapping spirals, and $\sim10\arcmin$ away from the closest of the two brightest cluster galaxies (BCGs) in Hydra\,I, i.e. NGC\,3311.

\begin{figure}[!htb]
   \centering
   \includegraphics[width=0.48\textwidth]{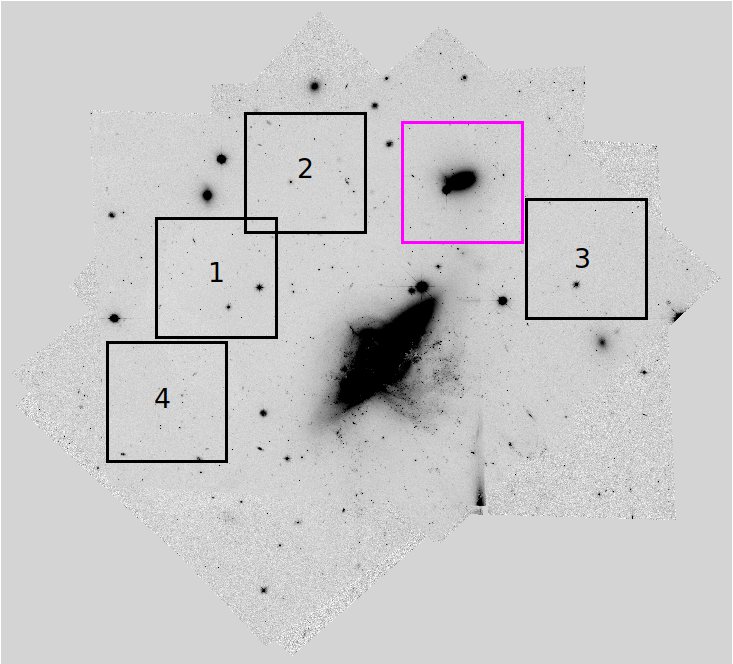}
   \caption{The image of the full frame in \g band with the region of \leda \, highlighted in magenta, and the four regions used as background frames in black. Each region is $1.1\arcmin \times 1.1\arcmin$. The most conspicuous object in the centre of this mosaic is NGC\,3314A/B.}
              \label{fig:images}%
    \end{figure}

Because of possible residual contamination of GCs belonging to the neighbouring BCG and spiral galaxies, as well as contamination from other fore- and back-ground sources, we compare the target frame with a set of background reference fields from the same HST/ACS mosaic. In particular, we used four background regions in the vicinity of \leda, chosen to be at approximately the same distance from NGC\,3314A/B, and far away from any obvious bright source in the field. The coordinates and properties of these fields are given in Table \ref{tab:subframes}. Figure \ref{fig:images} shows the position of \leda \, and the four background fields.

\begin{table}
\caption{Properties of the galaxy frame in each band}             
\label{tab:observations}      
\begin{tabular}{l l l }        
\hline \\ [-1.5ex]                
     Property  &  \g & \V \\  \\ [-1.5ex]   
\hline  \\ [-1.5ex]  
Exposure time (s) &62369.18&  64800 \\
Zeropoint (AB mag) & 26.071 & 26.507 \\
Encircled Energy (r=0.12")$^a$ & 0.7195 & 0.7145 \\
Extinction $A_{\lambda}$ (mag) & 0.207 & 0.157 \\
$FWHM$ & 0.11\arcsec &0.12\arcsec \\
\hline     \\              [-1.5ex]                
\end{tabular}
\footnotesize{\\Note: $a)$ See Sect. \ref{sec:photometry} for definition and details.}

\end{table}

\subsection{Modelling the galaxy} \label{sec:model}
In order to improve GC detection in regions of the galaxy with high surface brightness, we subtracted the galaxy mean profile. We modelled the galaxy with elliptical isophotes using the Astropy affiliated package photutils (v1.0.2) in Python.

We first initialized a rough galaxy model using first-guess ellipticity and position angle parameters using the EllipseGeometry class in photutils. Then we initialized an object of the Ellipse class with the unmasked galaxy image data and this geometry, and performed a short and coarse fit using the fit\_image routine within Ellipse. This provided us with a list of isophotes in the form of an object of the Isophote class, which were used to generate more refined starting parameters for the final fitting which would be performed on the masked image.

To obtain the mask, we ran the sewpy wrapper\footnote{See \url{https://github.com/megalut/sewpy}} for SExtractor \citep{bertin96}, generating separate photometric catalogs of the extended and compact objects in the frame. At this stage of the selection, we only masked  the brightest objects ($mag<24/25$ for extended/compact sources, respectively, in both passbands), and separated extended from compact sources using the SExtractor CLASS\_STAR parameter (CLASS\_STAR$>$0.85 for compactness).

\begin{figure}[!htb]
   \centering
   \includegraphics[width=0.48\textwidth]{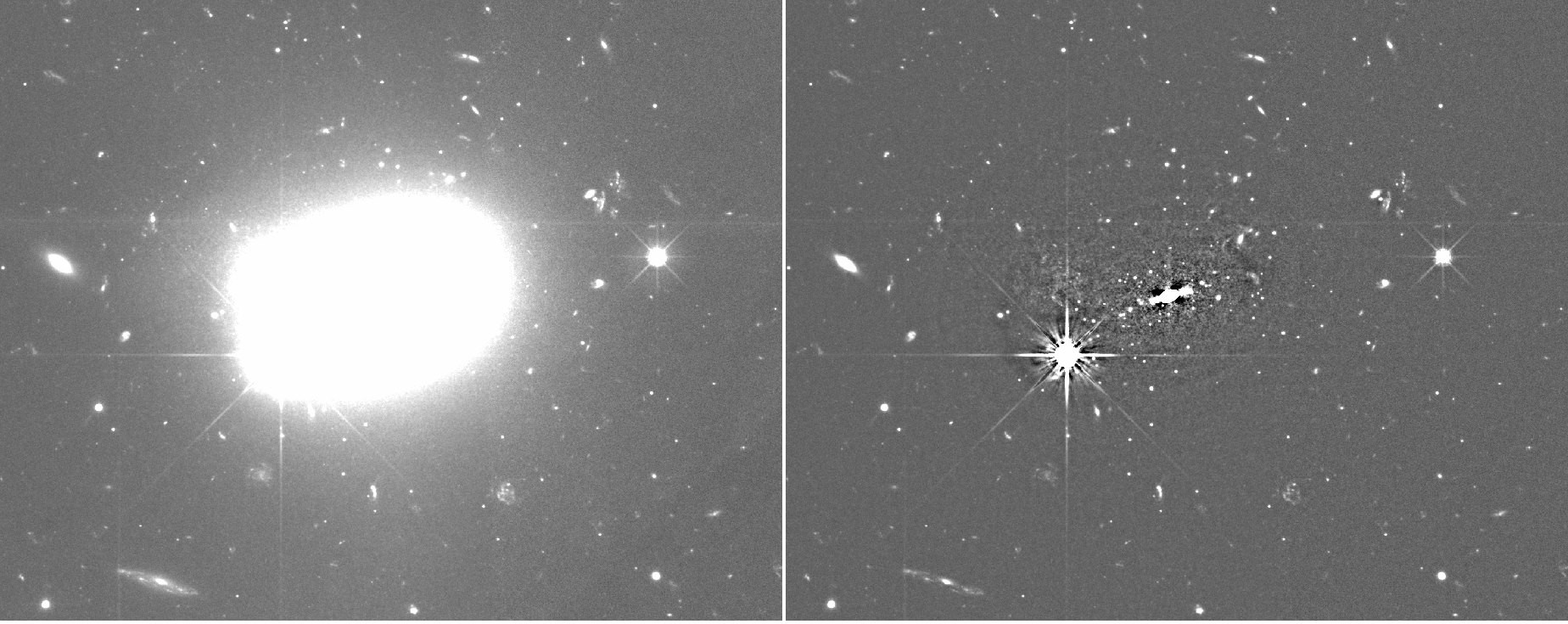}
   \caption{Left: \leda \, \g-band image. Right: galaxy subtracted residual image. We further subtracted a smooth background map derived from SExtractor, to improve the visibility of GC candidates near the galaxy center. The image size is $1.0\arcmin\times0.85\arcmin$}.
              \label{fig:residuals}%
    \end{figure}

The final fit for elliptical isophotes was then performed on the masked image of \leda~using the fit\_image routine. A low-surface brightness feature (possibly a diffuse foreground galaxy) and the bright star south-east of the core of \leda, both visible in Figure \ref{fig:closeup}, were also masked out before running the isophotes fitting. The galaxy modelling run went out to a semi-major axis of $26.5\arcsec$, where the galaxy surface brightness level reaches $\approx$26.3/26.0 $mag/arcsec^2$ in \g an \V-band, respectively. The model was synthesized from the list of isophotes generated by fit\_image, using the build\_ellipse\_model function within Ellipse. Finally, the residuals were generated by subtracting the model from the original image of \leda. 
Figure \ref{fig:residuals} shows the \g-band image of the galaxy and the smoothed residuals.

The Isophote class also provides us with information on the total flux within each fitted elliptical isophote, which is stored in the variable "$tflux\_e$". Using a curve of growth analysis on the total flux within each isophote, we determine the total apparent magnitude of the galaxy in both \g and \V. The values of this analysis are quoted in Tab. \ref{tab:properties}.

We carried out an additional test with Galfit \citep[][]{galfit} in order to estimate the disk-to-bulge ratio using a combined fit of an elliptical and a Sersic component to the galaxy. We find that the $B/T$ (bulge-to-total) ratio is $0.76$, implying an $D/B\approx 0.32$ (disk-to-bulge ratio). This, combined with the Sersic index of 5.8 from our Galfit run puts \leda \, firmly in the E/S0 category \citep[][]{baillard11}.

\subsection{Source detection and photometry}\label{sec:photometry}
Once we have the residual image, we re-run SExtractor to generate the catalog of sources in the image cutout. The catalog will be used to identify GCs in the frame. 

The photometry of point-like sources is aperture and extinction corrected as follows. We obtain the extinction correction from the IRSA Dust query module of Astropy, which gives us the value of $E_{B-V}$ at the position of our field(s). The value of $R_V$ is adopted from \citet{sf11}. 
The aperture correction values were obtained from the instrument web-pages, using the on-line encircled energy plots\footnote{See \url{https://www.stsci.edu/hst/instrumentation/acs/data-analysis/aperture-corrections}, derived by \citet{bohlin16}}, where the encircled energy is the fraction of flux contained within a certain radius of aperture \citep[][]{sirianni05}. We chose an aperture radius of $r=0.12\arcsec$, which corresponds approximately to the FWHM of our data (see Tab. \ref{tab:observations}). 

We verified the adopted encircled energy values by obtaining the aperture correction from the analysis of the most isolated, compact and bright objects in the field. By inspecting the curve of growth (i.e. the aperture magnitude at different radii) out to 64 pixels aperture, we obtained aperture correction values that agree within $\pm$0.04 mag with the ones from the HST/ACS calibration team. We finally matched the aperture and extinction corrected catalogs across the \g and \V-bands, using a matching radius of 3 pixels ($\sim0.12\arcsec$). This was done in order to select only the sources that appear in both bands.

The full catalog of GC candidates identified is available  Tab. \ref{tab:gccat}. In the catalog we provide the positions (cols. 2 and 3), the aperture and extinction corrected magnitudes with errors for the $g_{F475W}$ and $V_{F606W}$ bands (cols. 4-7), the \g-\V color, as well as the flux radius (half-light radius from SExtractor) $F_{rad}$ and $FWHM$ in each band (cols. 9-12) for the sample of selected GCs.

The same procedure for the photometry, extinction and aperture corrections, and cross-matching was also applied to the four background fields.

\begin{table}
\caption{Comparison of \leda \, and the background frames. The RA and Dec are the coordinates of the centre of each cutout. $N_{unsel}$ is the total number of sources, and $N_{GC}$ is the number of GC candidates detected in each frame.}    
\centering
\label{tab:subframes}  
\setlength{\tabcolsep}{3pt}
\begin{tabular}{l c c c c}        
\hline \\ [-1.5ex]                
Frame & RA & Dec & $N_{unsel}$ & $N_{GC}$ \\
 & (deg) & (deg) &  &\\ \\ [-1.5ex]   
\hline  \\ [-1.5ex]  
   \leda &159.290 &-27.658 &  804 & 102\\
    Ref 1 & 159.331 & -27.672& 485& 20 \\
    Ref 2  &159.316 &-27.657&  485 & 20 \\
    Ref 3  & 159.269 & -27.669& 369& 17 \\
    Ref 4  & 159.339 & -27.690& 449 & 20  \\ \\ [-1.5ex]
\hline                                   
\end{tabular}
\end{table}
\section{Globular Cluster population: sample selection}
\label{sec:gc_sel}
In this section we describe the procedures adopted to identify GC candidates from the cross-matched catalog derived in the previous section.

Since GCs at the distance of the Hydra\,I appear as point-like objects ($1\arcsec$ corresponds to nearly 100 pc), it is reasonable to identify them by means of their shape, in addition to their photometric properties. We selected GCs using three criteria based on $i$) compactness, $ii$) color and $iii$) magnitude.

The same selection procedure is applied to the catalogs of the four background fields.

\subsection{Compactness}
The selection on compactness is used to separate the compact GC candidates from the extended sources. We identify compact sources using the SExtractor $CLASS\_STAR$ parameter and the concentration index (CI).

For the SExtractor star-galaxy separation parameter $CLASS\_STAR$, which has a value close to 1.0 for compact objects, we adopted a threshold value of $CLASS\_STAR\geq0.8$.

The CI is defined as the difference between magnitudes at two different apertures, and is a further indicator of source compactness \citep[][]{peng11}. After several tests, we set the two aperture diameters at 4 and 8 pixels to calculate $CI=mag_{4pix}-mag_{8pix}$. We then selected all sources with $0.25\leq CI \leq 0.75$ in both passbands, based on the median and scatter of the sequence of compact sources in the CI vs magnitude plot.

\begin{figure}[!htb]
   \centering
   \includegraphics[width=0.48\textwidth]{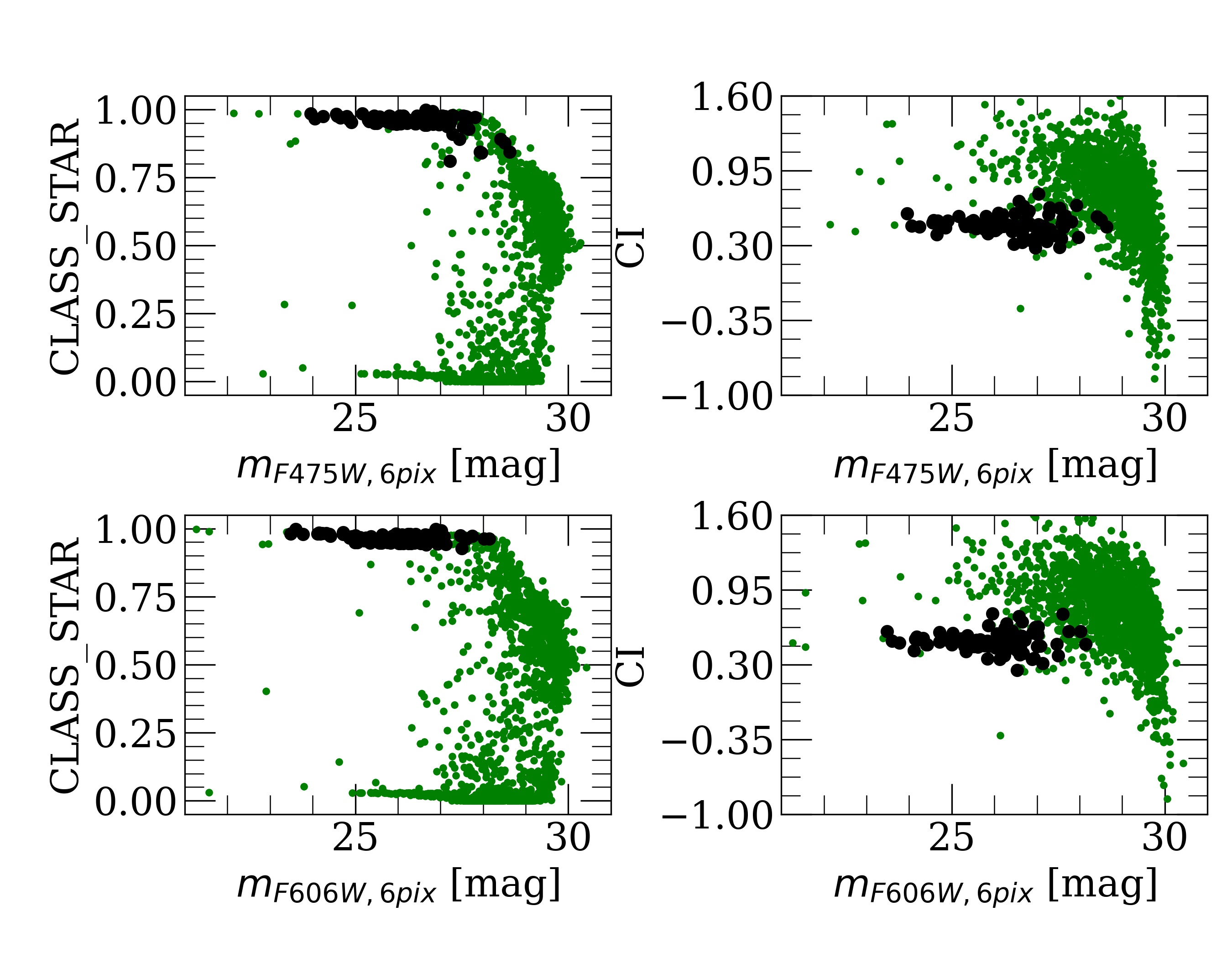}
   \caption{Left: Star-galaxy classifier of all detected sources in the matched catalog (green), and GC candidates (black). Right: Concentration indices for all the detected sources in the matched catalog (green) and GC candidates (black).}
               \label{fig:class_ci}%
    \end{figure}

The selection of compact sources is shown in Figure \ref{fig:class_ci} with black points.

\subsection{Color and magnitudes}\label{sec:colmag}
The selection based on color enables us to reduce the contamination from fore- and back-ground compact sources in the field, mostly Milky Way stars and distant galaxies, respectively.

\begin{figure*}[!htb]
   \centering
   \includegraphics[width=0.95\textwidth]{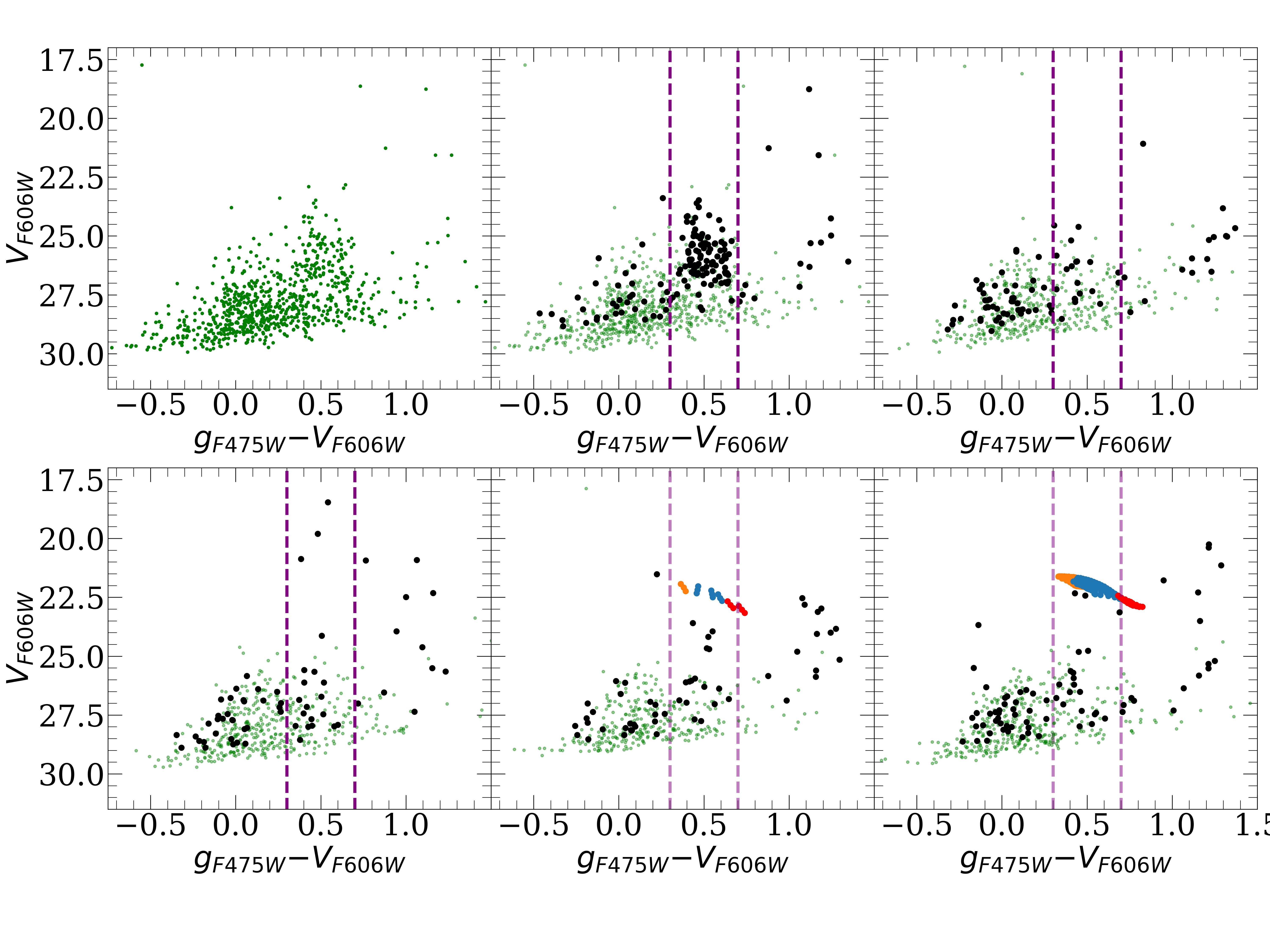}
   \caption{The color-magnitude diagram of all objects detected (green) (top left), and along with the objects selected by compactness criteria (black) in the \leda \,  frame (top centre), and background reference frames (all others). The GC candidates are the compact objects (black) within the color interval marked by purple dashed vertical lines, in all panels. The bottom two panels also show the SPoT (bottom left panel) and YEPS (bottom right panel) SSP models (with an arbitrary magnitude shift) overlaid on top of the color-magnitude plots for regions 3 and 4. Different SSP [Fe/H] are shown with different colors : $-2.5\leq[Fe/H]\leq-1.5$ in orange, $-1.5<[Fe/H]\leq0$ in blue, and $0<[Fe/H]\leq0.5$ in red. The vertical dotted lines in the middle panel show the color interval adopted for GCs selection.}
              \label{fig:colmag}%
\end{figure*}

Figure \ref{fig:colmag} shows the color-magnitude diagrams of sources detected in \leda \, and in the four reference fields. The upper left panel of the figure shows the color-magnitude diagram of the entire sample of matched sources with no selection. The panel reveals the presence of a sharp drop in the number of detected sources for \g-\V color bluer than $\sim$0.3 and redder than $\sim$0.7 mag, which is not seen in the four background fields, shown in the other panels of the figure.

Intermediate size galaxies like \leda \, typically do not contain a substantial population of red, metal-rich GCs \citep{peng06}. Using an updated version of the SPoT \citep[e.g.][]{raimondo05} simple stellar population (SSP) models, we estimate that with $[Fe/H]\leq0$, and age$\geq10$ Gyr, the range of color we expect for GCs is conservatively $0.3\leq \langle g_{F475W}{-}V_{F606W} \rangle\leq 0.7$ mag. We use this color range from SSP models, combined with the properties observed in the colour-magnitude diagrams in Fig. \ref{fig:colmag}, to select the GC candidates in our catalog. A similar range of GC colors was corroborated by examining the YEPS \citep[][]{yonsei} SSP models\footnote{The YEPS models can be found on http://cosmic.yonsei.ac.kr/YEPS.htm}. The SPoT models are reported in Tab. \ref{tab:spot} in the Appendix.

After the compactness and color selections over the matched catalogs, we also applied a cut on the magnitude. As anticipated in Sect. \ref{sec:intro}, the GCLF has a universal Gaussian shape, and a width $\sigma^{GCLF}$ which scales with the galaxy luminosity. Using eq. (5) from \citet{villegas10} adopting a total galaxy magnitude $M_{z,gal}\sim-20.4$ mag (see Tab. \ref{tab:properties}), we evaluated $\sigma_{F475W}^{GCLF}=0.93\pm0.04$ mag. Adopting a preliminary value of $M_{F475W}^{TO}\sim-7.3$ mag  (we will refine this in Sect. \ref{sec:results}), and a distance of 61 Mpc, we estimated the expected GCLF peak $m_{F475W}^{TOM}\sim26.6$ mag. We selected from the matched catalog all sources $\pm5\sigma_{F475W}^{GCLF}$ around the $m_{F475W}^{TOM}$, using a large interval in order to avoid introducing bias in our own distance estimate.

A synthesis of the GC candidates identified in \leda \, and the background fields is also given in Tab. \ref{tab:subframes}. In Figure \ref{fig:colmag} the sources identified as GC candidates are highlighted in black. The plots and data in Tab. \ref{tab:subframes} show that the galaxy hosts a factor of $\sim5$ more GCs than the background fields. From Figure \ref{fig:colmag}, we can verify that the number of objects selected through the compactness criteria with colors bluer than the GCs color, $\langle g_{F475W}{-}V_{F606W} \rangle<0.3$, have a similar density in the \leda~frame ($N_{red,gal}=35$) as in the background frames (median $N_{red,bkg}=35\pm9$). A similar trend is also observed in the population of objects redder than the GCs color, $\langle g_{F475W}{-}V_{F606W} \rangle>0.7$ ($N_{blue,gal}=15$, median $N_{blue,bkg}=14\pm2$). The last two panels in the figure also show the colors and magnitudes from the SPoT and YEPS SSP models, with a magnitude offset factored to match the magnitude range in the plots.

In summary, even though the GC selection is based on a single color and shape criteria, the plots in Figure \ref{fig:colmag} reveal a significant over-density of GC candidates in the galaxy frame compared to the background fields. In the forthcoming section we will take advantage of that to characterize the GC population.

Some of the objects in the GC candidates catalog have a larger than average flux radius, with SExtractror half-light radii $F_{rad}\gsim 2~pix$ (compared to the median $F_{rad}=1.7$ pix), relatively red colors ($\langle$\g-\V$\rangle$>0.53 mag) and magnitude $M_{F606W}\approx -7.1$ mag which make them probable candidates for "faint fuzzies", which have historically been observed with similar characteristics close to many lenticular galaxies \citep[][]{brodie06}.

\section{Globular Cluster population analysis} \label{sec:analysis}
\subsection{Color Distribution}

\begin{figure}[!htb]
   \centering
\includegraphics[width=0.48\textwidth]{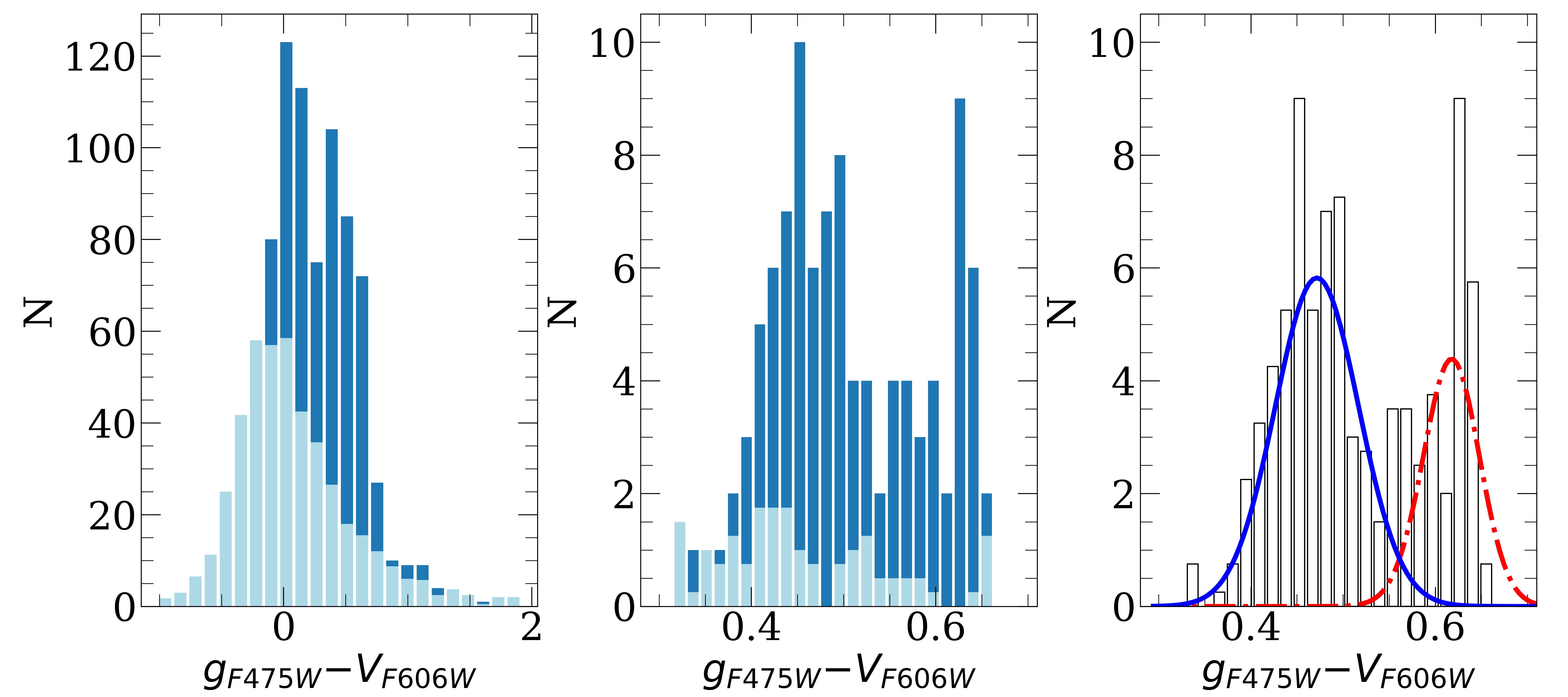}
\caption{Color histograms  of the full matched  (left panel), of GC candidates (centre) and the color distribution of GCs over \leda \, corrected for the background contamination (right).
The light blue histograms in the left and central panels indicate mean color distributions in the background frames, and the dark blue histograms represent the distribution in the frame of \leda. In the right panel the gaussian fits (from GMM) for the blue (solid blue line) and red (dot-dashed red line) GC sub-populations are overlaid on the background-corrected color density histogram.}
              \label{fig:colhist}%
\end{figure}
    
Despite the small separation in wavelength between the \g and \V passbands, the color histogram of the selected GCs shows a clear bimodality. By inspecting the entire population of matched sources on \leda \, and in the reference fields, it is hard to identify any evidence of this feature (Fig. \ref{fig:colhist}, left panel). However, the bimodality emerges when the color distribution of GC candidates is inspected, and it is especially evident when we subtract the background obtained from the four reference frames (Fig. \ref{fig:colhist}, middle and right panels).

To study the detailed properties of the color distribution, we generated a random sample of $\sim$1000 points in the shape of the background-subtracted histogram of GCs (Fig. \ref{fig:colhist}, right panel), using the $numpy$ routine random.choice. In order to make this distribution more continuous, each bin in this histogram of random points was smoothed with a uniform distribution having a width equal to half of the bin size. The resulting distribution, exhibiting a dual Gaussian profile, was fitted using Gaussian Mixture Models \citep[GMM,][]{muratov10}, repeated over 10 iterations. We found that a bimodal distribution is favored over a unimodal one, and the median blue (red) peak of the background subtracted density distribution lies at $\langle g_{F475W}{-}V_{F606W} \rangle=0.47~(0.62)$ mag, with a width of $0.05~(0.03)$ mag. The resulting fraction of red GCs is $f_{red}\sim0.3\pm0.1$, which agrees with the median value of $f_{red}\sim 0.18\pm0.20$ from \citet[][ their Tab. 3]{peng08} for galaxies similar to \leda, with a specific frequency ($S_N$) between 1.0 and 2.5 (further discussed in Sect. \ref{sec:results}).

Due to the lack of existing literature on the bimodality in \g-\V color, we projected the \g -\V peaks to $V{-}I$ and $g{-}z$ colors, using the SpoT and Yonsei models for the transformation. Considering an interval of $\langle$\g -\V$\rangle \pm0.05$ mag around each peak and adopting predictions for ages $t\geq 10$ Gyr and metallicity $-2.5\leq [Fe/H]\leq 0$, we found that the blue \g-\V=0.47 mag peak projects into $V{-}I=0.55$ mag and $g{-}z$=0.93 mag, while the red \g-\V=0.62 mag peak is projected into $V{-}I$=0.78 mag and $g{-}z$=1.34 mag. 
We compared the projected colors for each peak with the values from the review by \citet[][their Tab. 1]{brodie06}, for galaxies similar to \leda, having type S0 and $-19.5\leq M_B\leq -17.5$ mag. The median colors from this selection are $\langle V{-}I\rangle=0.51~(0.72)$ mag and $\langle g{-}z\rangle = 0.92~(1.26)$ mag for the blue (red) peak, which agree with our results within $\sim$0.1 mag. 
In summary, despite the narrow wavelength interval available, the GCs color bimodality appears to be consistent with the results obtained for similar galaxies over a wider wavelength range.

\subsection{Radial distribution}
\begin{figure*}[!htb]
   \centering
   \includegraphics[width=\textwidth]{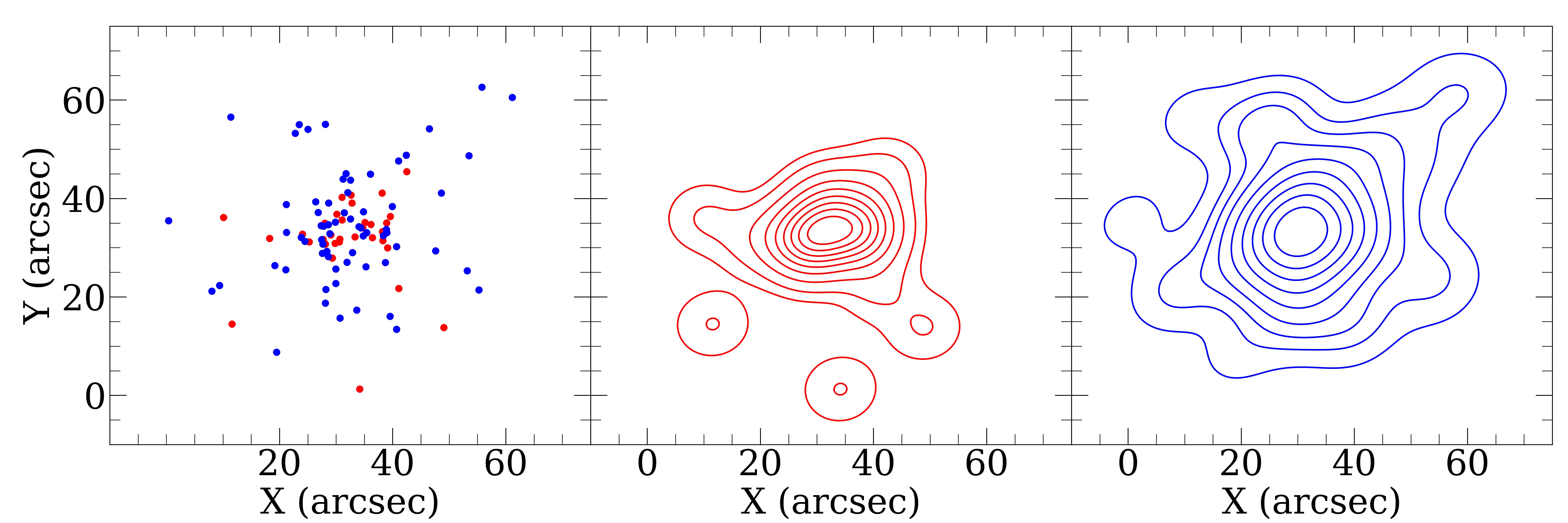}
   \caption{The spatial distribution of the GC candidates divided into red and blue according to color ($\langle g_{F475W}{-}V_{F606W} \rangle \geq0.53$ is red) in the frame of \leda \, (left) and the kernel density estimate(KDE) plot of the blue (centre) and red (right) GC sub-populations.} 
              \label{fig:spatial}%
    \end{figure*}

\begin{figure}[!htb]
    \centering
   \includegraphics[width=0.48\textwidth]{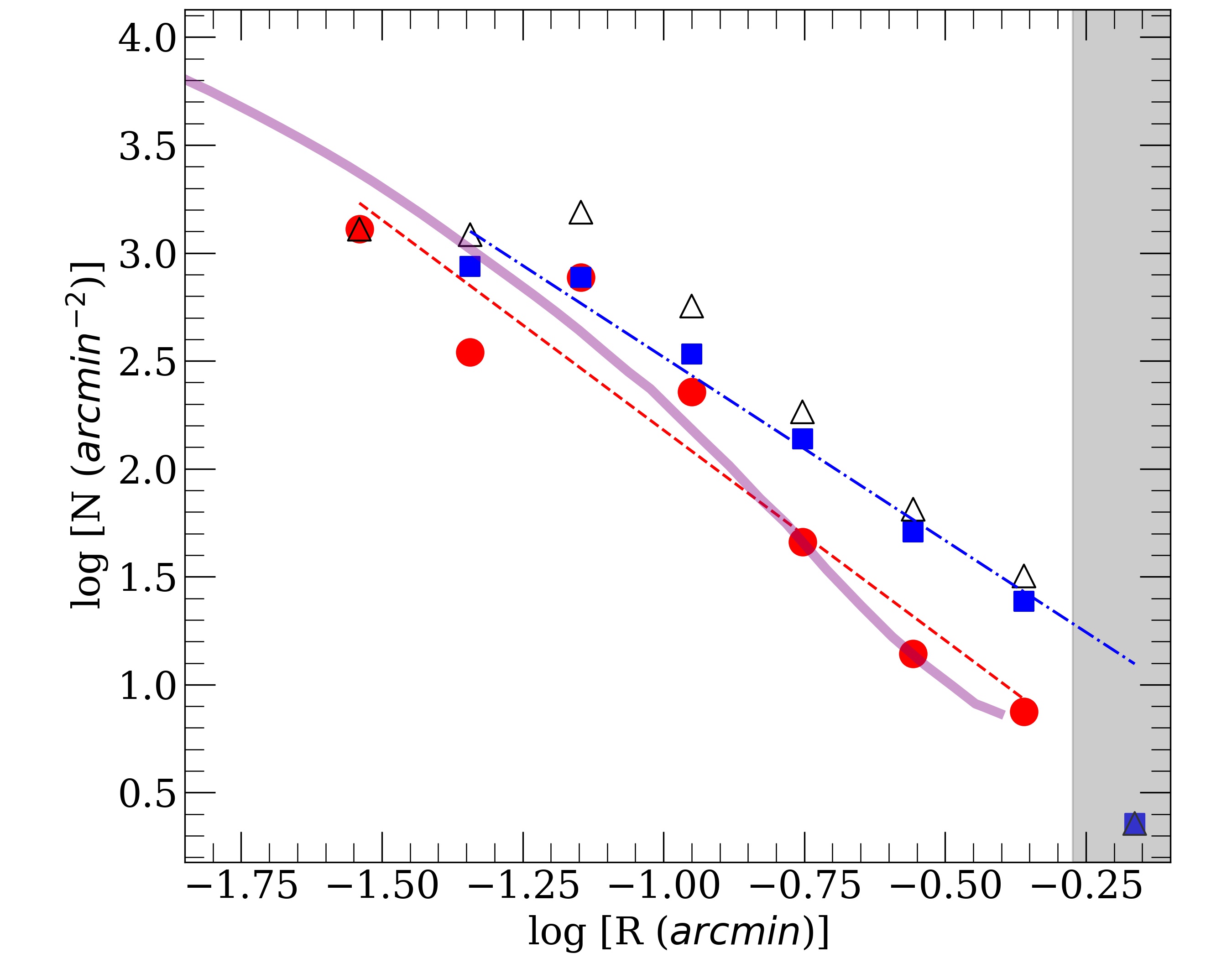}
    \caption{Radial densities of the full (black triangles), red (red circles) and blue (blue squares) GC populations in \leda.The radial density profile for the red (dashed line) and blue (dot-dashed line) GC populations is also shown as a linear trend. The radial surface brightness profile of the galaxy (scaled appropriately) is shown with the purple solid line. The shaded gray region shows the limit of geometric completeness as we approach the margins of our image cutout.}
    \label{fig:radlog}
\end{figure}

The coordinates of the GCs identified in each frame were plotted to understand the radial distribution of the sources in the frame of \leda \,(Fig. \ref{fig:spatial}). The GCs were separated into two different classes based on their $\langle g_{F475W}{-}V_{F606W}\rangle$ color: red ($\langle g_{F475W}{-}V_{F606W}\rangle\geq0.53$, adopted from GMM fits) and blue ($\langle g_{F475W}{-}V_{F606W}\rangle<0.53$). From Fig. \ref{fig:spatial} (left panel) and Fig. \ref{fig:radlog} we can identify that red, more metal-rich GCs appear clustered near the centre of the galaxy compared to the blue. The slope of the radial number density of the red GCs is steeper ($\alpha_r = \frac{d \log N^{red}_{GC}(r)}{d \log R} \approx-1.9$) than the blue ($\alpha_b\approx-1.7$). The slopes agree with the values of the slopes for metal-rich and metal-poor subpopulations quoted by \citet[][]{brodie06}. We also plotted the galaxy surface brightness radial profile (scaled appropriately), to emphasize its resemblance to the radial density profile of the red GCs. Around a radius of $\sim32\arcsec$ we begin to approach the edges of the galaxy cutout, and to avoid issues due to vignetting we do not consider this region for our fits (indicated by gray shaded area in Fig. \ref{fig:radlog}). In Fig. \ref{fig:spatial} we have plotted the kernel density estimate plots for the red and blue sub-populations, and we calculated that the ellipticity of the red GCs (centre frame) is $\epsilon =0.41$, where $\epsilon$ is the ratio of the minor to the major axis. The $\epsilon$ for red GCs matches that of our model isophotes in Sect. \ref{sec:model}. The KDE plot of the blue sub-population, on the other hand, has a more circular distribution with $\epsilon=0.13$ and a higher variation at larger radii.  

\subsection{Luminosity function}
\label{sec:gclf}
\begin{figure*}[!htb]
   \centering
      \includegraphics[width=\textwidth]{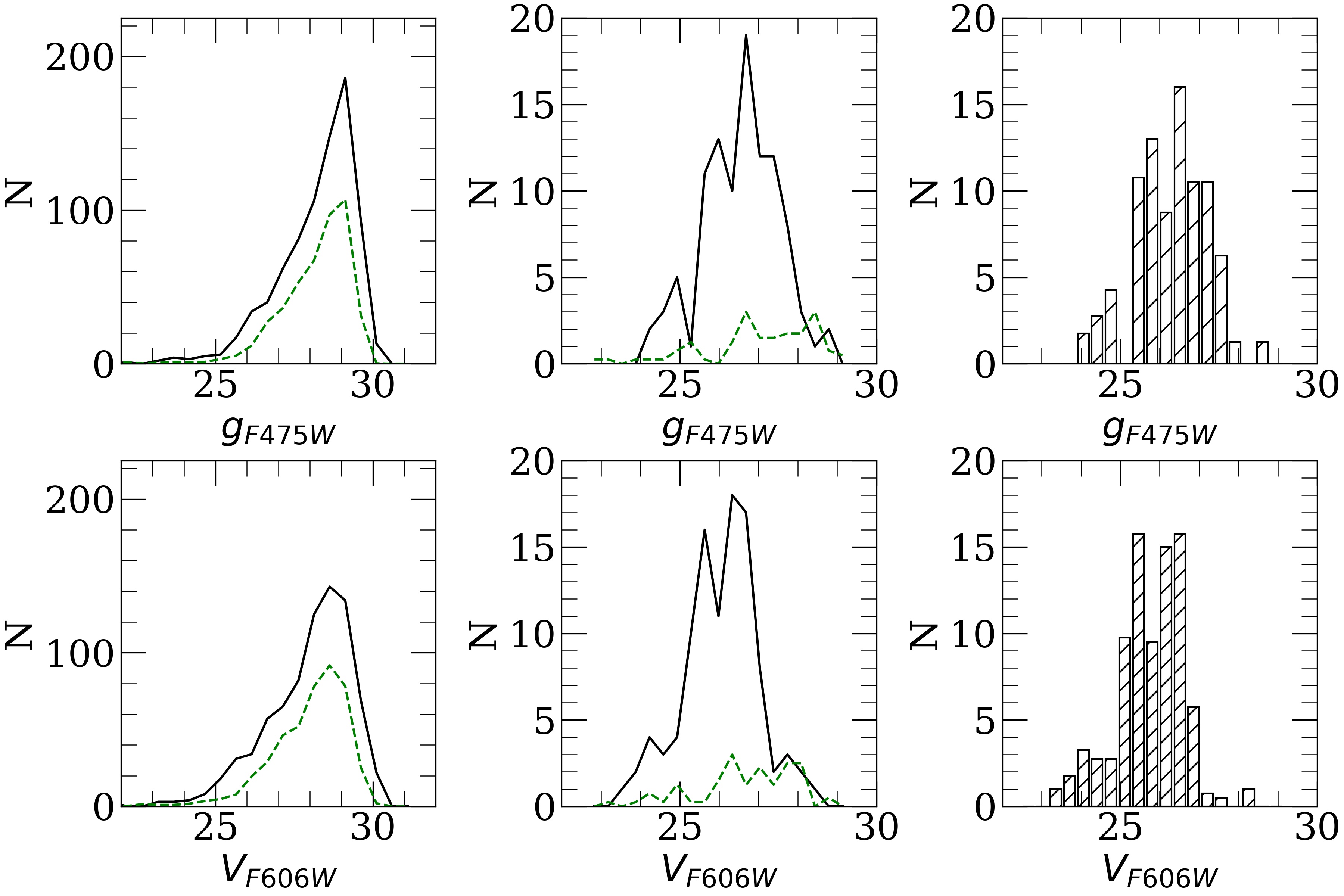}

   \caption{Left: The luminosity function (LF) of all detected and matched objects  in the frame of \leda \,(in black), and in the background fields (in green, dashed). Centre: The LF of the globular clusters candidates in the galaxy frame (in black), and the mean LF of GC candidates over the four background fields (in green). Right: The LF of the GC candidates in \leda \,after subtracting the mean LF from the background fields. This luminosity function has not yet been corrected for completeness.}
              \label{fig:results_gclf}%
    \end{figure*}

The histograms of GCs in the frame of \leda~and the background frames were inspected to study the luminosity function. As for the color and radial distribution, we used the four reference fields to define the background level of contaminant sources to be subtracted from the GC density over \leda. Since the area of the background fields is the same as the cutout of \leda, to properly subtract the background contamination it is not necessary to normalize to the area. Figure \ref{fig:results_gclf} shows the luminosity function of the sources detected in \leda~and (the mean) of the reference regions for the full matched catalogs. The middle panel of the same figure shows the luminosity functions for the GC candidates. As expected, the GCs on-galaxy significantly outnumber the counts in the reference fields. We also note a slight increase of background counts toward fainter magnitudes, as an effect of the lower efficiency of the adopted selection criteria for sources with lower S/N. The background subtracted luminosity function of GCs is shown in  Fig. \ref{fig:results_gclf} (right panel). In order to use the GCLF for deriving the distance modulus of \leda, it must be corrected for completeness first.

\begin{figure*}[!htb]
   \centering
   \includegraphics[width=0.4\textwidth]{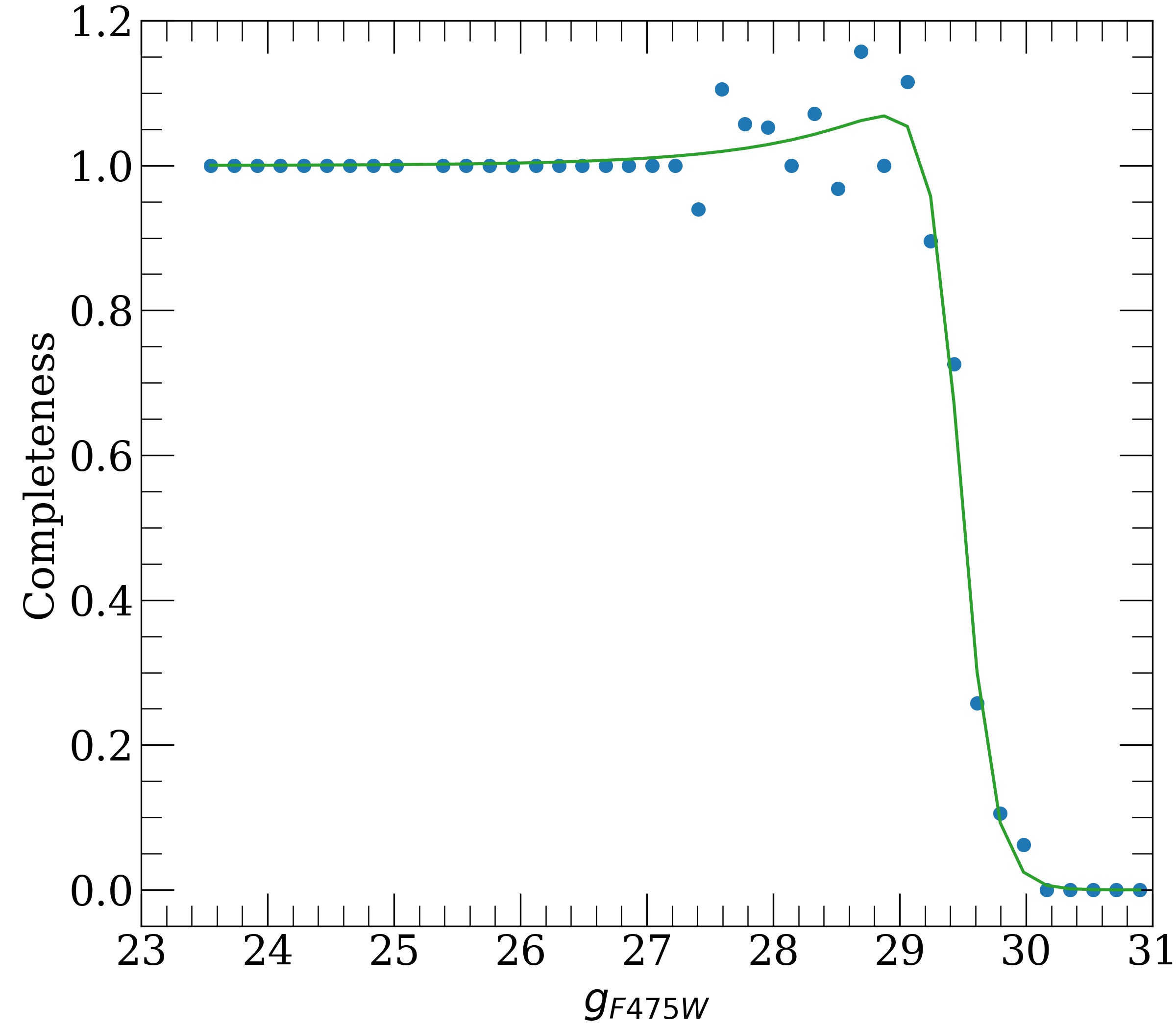}
   \includegraphics[width=0.4\textwidth]{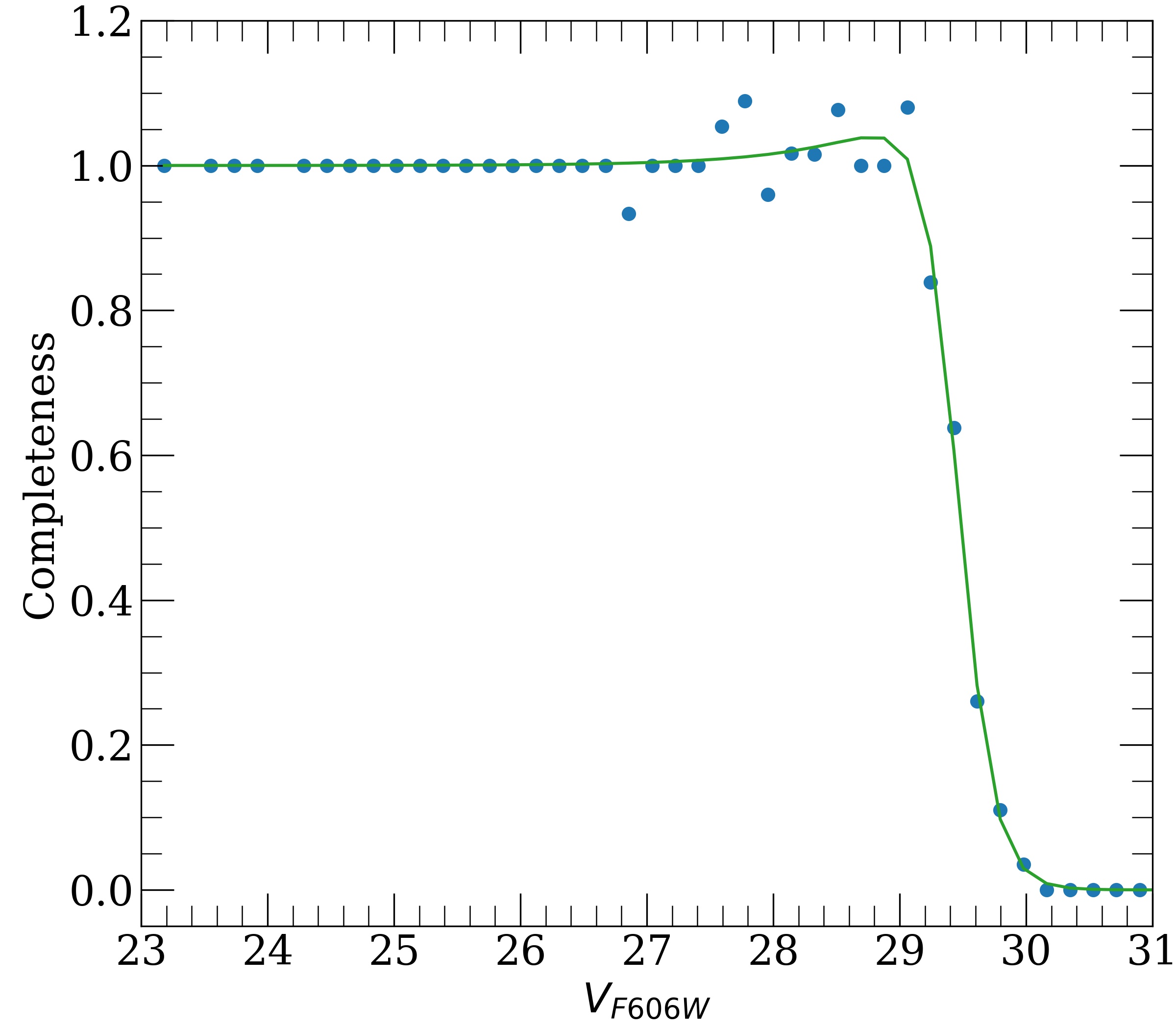}
   \caption{The completeness functions in \g (left) and \V (right). Each point represents the ratio of the number of sources detected vs  injected, in each magnitude bin, and the green line shows the best fit modified Fermi function.}
              \label{fig:completeness}%
    \end{figure*}

The completeness function in each band was determined by injecting and then detecting a total of $\sim850$ artificial stars on the \g and \V residual frames. At first, we derived a PSF from the bright, compact and isolated sources in the image in each band. The brightest and most isolated compact sources were extracted from the residual image using the extract\_stars routine of Astropy, and these extracted stars were then fed into the EPSFBuilder routine of the photutils package to generate an effective PSF of the size $41\times41$ pixels. The magnitudes of artificial stars to be injected were obtained by generating a random sample of magnitudes, again using numpy.random.choice, in the shape of the histogram of all sources detected in the galaxy frame, limited to the range $23\leq mag\leq 32$. These stars were injected in the residual frame of \leda \,along an equispaced grid whose position was varied over 50 iterations. A catalog of detected sources was obtained from a run of SExtractor on this synthesized image, adopting the same parameters used for generating the GCs catalog. The aperture correction for each band was then applied on this catalog. A two-fold selection algorithm was used to match the detected to the injected artificial stars: a first selection on separation ($\leq 6$ pix), followed by a cut on the magnitude difference (absolute difference $|m_{injected}-m_{detected}|\leq 0.5$ mag). The ratio of the number of sources thus retrieved vs injected, in each magnitude bin, gives us the discrete completeness function (Fig. \ref{fig:completeness}).

We fit the completeness function with a modified Fermi function \citep[][their eq 2]{alamo13}, and interpolated it to counteract the discrete nature of the sampled magnitudes, then applied it to correct the GCLF in the galaxy frame. We note that the completeness is $90\%$ down to about $m\approx29.5$ in both \g and \V, which is fainter than the 3$\sigma_{GCLF}$ tail of the galaxy $m_{TO}$. In other words, the observed GCLF is complete and only mildly affected by the completeness, and a correction $\sim5\%$ is required for magnitudes \g$=28-29$ mag. The same analysis was performed on the background frames to determine their completeness functions individually, and the mean of the completeness-corrected LFs of GC candidates from the four background frames was used to determine and subtract the overall background for the GCLF in \leda.

\begin{figure}[!htb]
   \centering
      \includegraphics[width=0.45\textwidth]{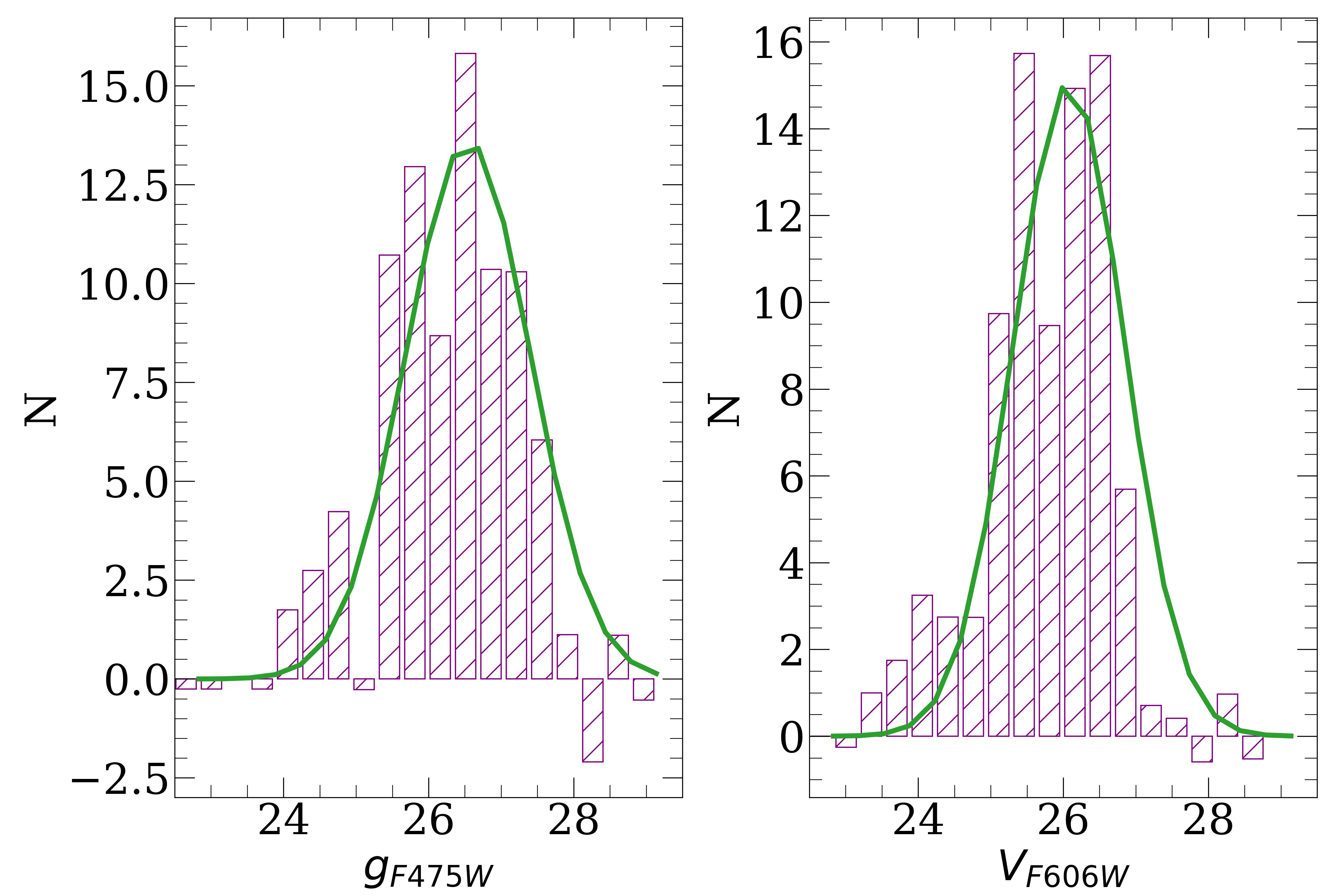}
   \caption{The final GCLF which has been corrected for completeness, and fitted with a Gaussian (green line).}
       \label{fig:gclf_fit}

\end{figure}

Finally, the background-subtracted GCLF was fitted with a Gaussian in both the \g and \V bands (Fig. \ref{fig:gclf_fit}) to obtain the peak turn over apparent magnitude and the width, the results of which are given in Table \ref{tab:gclf}.

\begin{table}
\caption{GCLF parameters}             
\label{tab:gclf}  
\setlength{\tabcolsep}{3pt}
\centering                          
\begin{tabular}{ c c c c c}      
\hline \\ [-1.5ex]                
Band & $m_{TO}$ & $\sigma_{GCLF}$ & $M_{TO}$& $m-M$ \\   
\hline  \\ [-1.5ex]  
    \g  & 26.54\textpm 0.10 & 0.86\textpm 0.10 &-7.29\textpm 0.28&33.76\textpm 0.30 \\
    \V & 26.08\textpm 0.09 & 0.76\textpm 0.09&-7.69\textpm 0.18&33.77\textpm 0.20 \\ \\ [-1.5ex]
\hline                                   
\end{tabular}
\end{table}

\subsection{GCLF distance: calibration and results} \label{sec:results}

In order to derive the GCLF distance to \leda \,we adopt the apparent $m_{TO}$ derived as described in Sect \ref{sec:gclf} combined with $M_{TO}$ calibrations in \g and \V from the literature, outlined below. 

We estimated the absolute \V band turn-over by taking advantage of existing accurate GCLF analysis by \citet[][]{iskrenpuzia09} on HST/ACS data of dwarf galaxies. We use their calibration quoted in the V-band, which gives us $M_V^{TO}=-7.60\pm0.11$ mag. Using the photometric transformation in Tab. 21 from \citet[][]{sirianni05} and inverting the relation given in eq. 12, we find the $M^{TO}_{F606W}$:
\[M_{F606W}^{TO}=M_V^{TO}-c_0-c_1\times(V-I) -c_2\times(V-I)^2+Zpt_{F606W}(AB)\]

The color used in this equation is the $V-I$ color of GCs, which is $\sim1$ \citep[][]{harris96,cantiello07c}, and $Zpt_{F606W}(AB)$ is the zero-point magnitude of the F606W band (Tab. \ref{tab:observations}) for the date of our observation, obtained from the zero-point calculator of the HST/ACS\footnote{Can be found at URL : https://acszeropoints.stsci.edu/}. Performing this analysis for the observed and synthetic (color$>0.4$) coefficients in \citet[][]{sirianni05} we find two individual estimates: $M^{TO}_{F606W,obs}=-7.65\pm0.14$ mag and $M^{TO}_{F606W,syn}=-7.73\pm0.11$ mag. On averaging these two values, we get $M^{TO}_{F606W}=-7.69\pm0.18$ mag.  

To obtain the \g band $M^{TO}$ we used the ACSVCS results by \citet[][their Tab. 1]{villegas10} for the Virgo cluster sample. We isolated a sub-sample of galaxies with $-19\leq M_B~(mag) \leq -17$ which corresponds to the magnitude level of \leda \,(see Tab. \ref{tab:properties}). From this sample, we excluded the galaxies with a number of GCs $N_{GC}<30$, as well as VCC\,1025 (since it belongs to the W' cloud and has a different distance modulus). We calculated the median turn-over magnitude of the selected sub-sample and the $rms$ (derived from the median absolute deviation), which turns out to be $m^{TO}_{F475W}=23.87\pm 0.25$ mag. Adopting $(m{-}M)_{SBF}=31.09\pm0.15$ for the Virgo cluster from \citet[][]{blake09}, we find :
\[M^{TO}_{F475W}=m^{TO}_{F475W}-(m{-}M)_{SBF}=-7.22\pm0.28\:mag\]

We calculate the distance modulus $m^{TO}{-}M^{TO}$ in each band, the results of which are in Tab. \ref{tab:gclf}. Being independent measurements (except the catalogs being cleaned by matching the \g and \V photometry) we assume our best distance modulus to \leda~to be the weighted average of the two values in Table \ref{tab:gclf}: $(m{-}M)_{GCLF}=33.77\pm0.17$ mag, or $56.7\pm 4.3$ Mpc. This agrees with the distance estimate from the Fundamental Plane for this galaxy, $(m-M)=33.8\pm0.5$ mag \citep[][]{tully13}. 

The systematic error on the calibrations derived above, could be assumed to be of the same order of the SBF systematic, i.e. $\sim0.1$ mag \citep{cantiello18}. \citet{lee18} claim that the real global systematic for GCLF is anyway $<0.3$ mag. We adopt a conservative middle-ground estimate of 0.2 mag as our systematic error budget. Hence, including systematic errors we obtain $D_{GCLF}=56.7\pm 4.3(statistical) \pm5.2 (systematic)$ Mpc.

We estimate the value of the Hubble parameter using the flow corrected radial velocity for \leda \, from Table \ref{tab:gclf}. Combining it with the distance modulus from this work we find $H_0\approx78.5\pm6.0(statistical)\pm7.3(systematic)$ \ksm.

The distance modulus from this work places \leda~ between NGC\,3314A \citep[$(m-M)=33.19\pm 0.40$ mag, from][]{theureau07} and NGC\,3314B \citep[$(m-M)=34.37\pm 0.15$ mag, from][]{mould07} in projection, consistent with other distance measurements for the Hydra cluster and lying close to the upper end of these estimates \citep[for e.g., ][]{blake02,hudson04,mieske05}. Our results show that this galaxy lies at a distance that is $\sim20\%$ larger than that of the two BGCs in the Hydra\,I cluster \citep[][]{tully13}, and in a region where we count eight other brighter galaxies which have upto 20\% larger velocity distances than that of \leda~in the vicinity of $\sim15\arcmin$\footnote{Source: Nasa Extragalatic Database}.

Considering the Hubble-Lemaitre distance to \leda~ and the results in Table \ref{tab:gclf} we can turn the argument around, and use our results to check whether the fitted parameters support the universality of the GCLF. 
Given the fact that the final distance calculated in our work agrees closely with the velocity distance (reported in Table \ref{tab:properties}), combined with the $\sigma_{GCLF}$ values from our analysis that match very well with the expected $\sigma_{GCLF}=0.93$ mag (see Sect. \ref{sec:colmag}), we can conclude that the results from this work support the universality of the GCLF method.

\subsection{Specific frequency}
We can also estimate the specific frequency \citep[][]{harris81} of the GC system in \leda, which is defined as:
\[S_N\equiv N_{GC}\times10^{0.4(M_{V,gal}+15)}\]

Using the $m_{F606W,gal}$ obtained from our fit in Sect. \ref{sec:model} (quoted in Tab. \ref{tab:properties}), correcting it for extinction and using Tab. 21 from \citet[][]{sirianni05} we transform $m_{F606W,gal}$ to $m_{V,gal}$ in Vegamag, which is used as reference for $S_N$ estimates.

Combining this apparent magnitude with the distance modulus of \leda~(from Sec. \ref{sec:results}), we estimate a specific frequency $S_N=1.8\pm0.7$. The value we estimated is consistent with the observed scatter of $S_N$ for galaxies of similar magnitude \citep[][]{peng08,harris13}, leaning towards the tail of higher GC population density.

\section{Conclusions}\label{sec:conclusions}

In this work, we benefited from extremely deep images in the HST/ACS \g and \V bands to characterize the GC population around \leda. The main results of our study are summarized below:
\begin{itemize}
    \item  Although the population of GCs we find is relatively small, $N_{GC}=82\pm9$, and the wavelengths of the \g and \V bands are relatively close, we find a clear bimodal color distribution in the GC system of this galaxy;
    \item The radial distribution of the GCs shows a clustering of the red GCs close to the core of the galaxy, which falls off rapidly at higher radii ($\alpha_r \approx-1.9$) and has a close resemblance to the galaxy surface brightness profile, whereas the blue GCs are less concentrated at the centre, taper off more slowly away from the galaxy core ($\alpha_b \approx-1.7$) and appear circularly distributed around the galaxy.
    \item In spite of the close wavelengths of the \g and \V bands, and the intermediate mass of \leda, we observe the typical bimodal characteristics of GC populations that are more pronounced in massive galaxies. The \g-\V color distribution shows a blue (red) peak at $\langle g_{F475W}{-}V_{F606W} \rangle=0.47(0.62)$. This further demonstrates the role of quality of the images on the characterization of GC populations, since a lower depth and accuracy of images would have smeared out the color bimodality feature we observe in this work;
    \item The turn-over magnitude in \g and \V are both $\sim3$ mag brighter than the completeness limit ($>29.5$ mag), which to our knowledge is an unprecedented finding at distances of the order of 60 Mpc; 
    \item We estimate a distance modulus of $(m{-}M)=33.77\pm0.17$ mag, or $56.7\pm 4.3(statistical)$ Mpc for this galaxy, which places it between NGC\,3314A and NGC\,3314B in projection;
    \item Considering the velocity distance of \leda ~and the empirical expectations for the GCLF peak and width, our analysis for this target supports the universality of the GCLF method;
    \item Assuming $\sim10\%$ systematic error on the galaxy distance, we derive a Hubble constant value of $H_0\approx78.5\pm6.0(statistical)\pm7.3(systematic)$ \ksm. 
\end{itemize}

\begin{acknowledgements}
This work was carried out based on observations made with the NASA/ESA Hubble Space Telescope, and obtained from the Hubble Legacy Archive, which is a collaboration between the Space Telescope Science Institute (STScI/NASA), the Space Telescope European Coordinating Facility (ST-ECF/ESA) and the Canadian Astronomy Data Centre (CADC/NRC/CSA). The authors made use of Astropy,\footnote{http://www.astropy.org} a community-developed core Python package for Astronomy \citep{astropy:2013, astropy:2018}, along with the databases of HyperLeda \citep[][]{makarov14}, the Extragalactic Distance Database (EDD)\footnote{https://edd.ifa.hawaii.edu/} and the NASA/IPAC Extragalactic Database (NED,
which is operated by the Jet Propulsion Laboratory, California Institute of Technology,
under contract with the National Aeronautics and Space Administration). We also made extensive use of the softwares of Topcat\footnote{http://www.starlink.ac.uk/topcat/}, SExtractor \citep[][]{bertin96} and Galfit \citep[][]{galfit}. MC acknowledges support from MIUR, PRIN 2017 (grant 20179ZF5KS). The authors would also like to acknowledge and thank the referee for their valuable comments, questions and suggestions.
\end{acknowledgements}

\bibliographystyle{aa}
\bibliography{nandinih_oct21} 

\begin{appendix}
\onecolumn
\section{Tables}
\begin{table}
\caption{SPoT models}
\label{tab:spot}
\begin{tabular}{cccccc}\\ [-1.5ex]   
\hline  \\ [-1.5ex]  
z      & Age & $M_V$        & $V$-\g   & $V$-\V  & \g-\V\\
       &(GYr)&(Vegamag)&(Vegamag)         &(Vegamag)    &(Vegamag)  \\ \\ [-1.5ex]   
\hline  \\ [-1.5ex]  
0.0001 & 2    & -12.8572 & -0.2743 & 0.1329 & 0.4072 \\
0.0001 & 3    & -12.5237 & -0.3051 & 0.144  & 0.4491 \\
0.0001 & 4    & -12.282  & -0.3301 & 0.1563 & 0.4864 \\
0.0001 & 6    & -11.9145 & -0.3521 & 0.1623 & 0.5144 \\
0.0001 & 7    & -11.79   & -0.3609 & 0.1662 & 0.5271 \\
0.0001 & 8    & -11.674  & -0.3645 & 0.1684 & 0.5329 \\
0.0001 & 10   & -11.4846 & -0.3741 & 0.1748 & 0.5489 \\
0.0001 & 12   & -11.3264 & -0.3856 & 0.1805 & 0.5661 \\
0.0001 & 14   & -11.173  & -0.3947 & 0.1833 & 0.578  \\
0.001  & 2    & -12.8318 & -0.3423 & 0.1602 & 0.5025 \\
0.001  & 3    & -12.4435 & -0.3725 & 0.1707 & 0.5432 \\
0.001  & 4    & -12.1836 & -0.4004 & 0.1843 & 0.5847 \\
0.001  & 6    & -11.7924 & -0.4258 & 0.1947 & 0.6205 \\
0.001  & 7    & -11.6625 & -0.4341 & 0.1979 & 0.632  \\
0.001  & 8    & -11.5549 & -0.4412 & 0.2009 & 0.6421 \\
0.001  & 10   & -11.363  & -0.4468 & 0.2034 & 0.6502 \\
0.001  & 12   & -11.2115 & -0.4433 & 0.2032 & 0.6465 \\
0.001  & 14   & -11.0623 & -0.4394 & 0.2031 & 0.6425 \\
0.004  & 2    & -12.638  & -0.3996 & 0.1844 & 0.584  \\
0.004  & 3    & -12.2876 & -0.4364 & 0.1998 & 0.6362 \\
0.004  & 4    & -11.9106 & -0.4526 & 0.2087 & 0.6613 \\
0.004  & 6    & -11.5842 & -0.4719 & 0.2139 & 0.6858 \\
0.004  & 7    & -11.4589 & -0.4823 & 0.2184 & 0.7007 \\
0.004  & 8    & -11.3427 & -0.4908 & 0.2221 & 0.7129 \\
0.004  & 10   & -11.1549 & -0.5015 & 0.2263 & 0.7278 \\
0.004  & 12   & -10.9981 & -0.5044 & 0.2286 & 0.733  \\
0.004  & 14   & -10.8679 & -0.5069 & 0.2301 & 0.737  \\
0.008  & 2    & -12.445  & -0.4307 & 0.1973 & 0.628  \\
0.008  & 3    & -12.1208 & -0.4667 & 0.2114 & 0.6781 \\
0.008  & 4    & -11.8348 & -0.4827 & 0.2176 & 0.7003 \\
0.008  & 6    & -11.4408 & -0.5024 & 0.225  & 0.7274 \\
0.008  & 7    & -11.2877 & -0.5094 & 0.2278 & 0.7372 \\
0.008  & 8    & -11.1762 & -0.5182 & 0.2313 & 0.7495 \\
0.008  & 10   & -10.982  & -0.5307 & 0.2364 & 0.7671 \\
0.008  & 12   & -10.8242 & -0.5397 & 0.24   & 0.7797 \\
0.008  & 14   & -10.6995 & -0.5473 & 0.2436 & 0.7909 \\
0.02   & 2    & -12.1286 & -0.4662 & 0.2079 & 0.6741 \\
0.02   & 3    & -11.7954 & -0.4984 & 0.2208 & 0.7192 \\
0.02   & 4    & -11.5515 & -0.5195 & 0.2295 & 0.749  \\
0.02   & 6    & -11.1586 & -0.5425 & 0.2389 & 0.7814 \\
0.02   & 7    & -10.9995 & -0.5499 & 0.2423 & 0.7922 \\
0.02   & 8    & -10.8795 & -0.5589 & 0.2461 & 0.805  \\
0.02   & 10   & -10.6701 & -0.572  & 0.2517 & 0.8237 \\
0.02   & 12   & -10.5088 & -0.5843 & 0.2569 & 0.8412 \\
0.02   & 14   & -10.3779 & -0.5954 & 0.2616 & 0.857  \\
0.04   & 2    & -11.8901 & -0.5012 & 0.2224 & 0.7236 \\
0.04   & 3    & -11.5415 & -0.535  & 0.236  & 0.771  \\
0.04   & 4    & -11.2996 & -0.5539 & 0.2441 & 0.798  \\
0.04   & 6    & -10.9501 & -0.5799 & 0.2553 & 0.8352 \\
0.04   & 7    & -10.8114 & -0.5933 & 0.2611 & 0.8544 \\
0.04   & 8    & -10.67   & -0.6002 & 0.2643 & 0.8645 \\
0.04   & 10   & -10.4599 & -0.617  & 0.2718 & 0.8888 \\
0.04   & 12   & -10.2896 & -0.6303 & 0.2778 & 0.9081 \\
0.04   & 14   & -10.1477 & -0.6416 & 0.2828 & 0.9244
\end{tabular}
\end{table}
\FloatBarrier

\begin{landscape}
\begin{longtable}{lllllllcllll}
\caption{GC catalog}
\label{tab:gccat}\\  

\hline 
Num & RA (J2000) & Dec (J2000)& $g_{F475W}$ & $\Delta g_{F475W}$  & $V_{F606W}$ & $\Delta V_{F606W}$&\g-\V&$F_{rad,F475W}$& $FWHM_{F475W}$&$F_{rad,F606W}$& $FWHM_{F606W}$\\
 & [deg]&[deg]&[mag]&[mag] & [mag]&[mag]&[mag] & [pix]&[pix]&[pix]&[pix]\\\\ [-1.5ex]
\hline  \\ [-1.5ex]
1      & 159.28939 & -27.66667 & 28.412 & 0.062       & 27.75  & 0.028       & 0.662 & 1.898     & 3.67    & 1.991     & 3.54    \\
2      & 159.294   & -27.66458 & 27.278 & 0.022       & 26.893 & 0.013       & 0.385 & 1.935     & 3.31    & 1.854     & 3.28    \\
3      & 159.29552 & -27.6633  & 26.68  & 0.052       & 26.234 & 0.022       & 0.446 & 1.798     & 3.47    & 1.982     & 3.5     \\
4      & 159.28733 & -27.66329 & 26.449 & 0.01        & 25.871 & 0.006       & 0.578 & 1.727     & 2.95    & 1.727     & 2.87    \\
5      & 159.28472 & -27.66319 & 27.623 & 0.008       & 27.081 & 0.004       & 0.542 & 1.618     & 2.83    & 1.709     & 2.78    \\
6      & 159.29647 & -27.66299 & 27.037 & 0.03        & 26.598 & 0.015       & 0.439 & 1.626     & 2.8     & 1.62      & 2.83    \\
7      & 159.29048 & -27.66266 & 26.412 & 0.018       & 25.997 & 0.01        & 0.415 & 1.737     & 2.71    & 1.743     & 2.9     \\
8      & 159.28771 & -27.66255 & 26.682 & 0.01        & 26.292 & 0.006       & 0.39  & 1.723     & 2.87    & 1.702     & 2.9     \\
9      & 159.28956 & -27.66221 & 26.13  & 0.013       & 25.664 & 0.008       & 0.466 & 2.252     & 3.56    & 2.104     & 3.58    \\
10     & 159.29129 & -27.66182 & 28.629 & 0.008       & 28.139 & 0.005       & 0.49  & 1.862     & 2.94    & 1.795     & 2.96    \\
11     & 159.28278 & -27.66107 & 26.782 & 0.01        & 26.421 & 0.006       & 0.361 & 1.87      & 3.53    & 1.952     & 3.24    \\
12     & 159.29126 & -27.66104 & 26.317 & 0.01        & 25.837 & 0.006       & 0.48  & 1.665     & 2.86    & 1.746     & 3       \\
13     & 159.28722 & -27.66098 & 26.653 & 0.01        & 26.102 & 0.006       & 0.551 & 1.663     & 2.72    & 1.847     & 2.95    \\
14     & 159.29716 & -27.66081 & 25.586 & 0.005       & 25.108 & 0.003       & 0.478 & 1.734     & 2.92    & 1.795     & 2.95    \\
15     & 159.29071 & -27.6607  & 25.76  & 0.006       & 25.329 & 0.004       & 0.431 & 1.716     & 2.73    & 1.718     & 2.85    \\
16     & 159.28341 & -27.65999 & 25.823 & 0.004       & 25.398 & 0.002       & 0.425 & 1.775     & 3.04    & 1.744     & 2.91    \\
17     & 159.29349 & -27.65992 & 26.654 & 0.014       & 26.185 & 0.009       & 0.469 & 1.566     & 2.85    & 1.652     & 2.94    \\
18     & 159.29071 & -27.65988 & 24.668 & 0.002       & 24.22  & 0.002       & 0.448 & 1.766     & 2.9     & 1.794     & 2.97    \\
19     & 159.28904 & -27.65975 & 27.808 & 0.042       & 27.467 & 0.03        & 0.341 & 1.637     & 3.49    & 1.619     & 2.98    \\
20     & 159.29409 & -27.65969 & 27.084 & 0.02        & 26.568 & 0.012       & 0.516 & 1.46      & 2.69    & 1.601     & 2.81    \\
21     & 159.28797 & -27.65952 & 26.467 & 0.011       & 26.046 & 0.007       & 0.421 & 1.735     & 2.93    & 1.795     & 3.23    \\
22     & 159.29009 & -27.6595  & 25.81  & 0.008       & 25.3   & 0.005       & 0.51  & 1.689     & 2.69    & 1.702     & 2.97    \\
23     & 159.2909  & -27.65927 & 26.673 & 0.019       & 26.07  & 0.012       & 0.603 & 1.542     & 2.91    & 1.601     & 2.93    \\
24     & 159.29113 & -27.65919 & 26.825 & 0.023       & 26.311 & 0.017       & 0.514 & 1.321     & 2.72    & 1.821     & 4.03    \\
25     & 159.28978 & -27.65895 & 27.965 & 0.009       & 27.497 & 0.007       & 0.468 & 1.595     & 2.76    & 1.68      & 2.85    \\
26     & 159.2912  & -27.65889 & 25.811 & 0.022       & 25.315 & 0.016       & 0.496 & 1.687     & 2.88    & 1.989     & 3.93    \\
27     & 159.28517 & -27.65886 & 26.575 & 0.003       & 26.06  & 0.002       & 0.515 & 1.758     & 2.87    & 1.786     & 2.93    \\
28     & 159.28785 & -27.65868 & 25.409 & 0.021       & 24.919 & 0.013       & 0.49  & 1.628     & 2.93    & 1.629     & 2.87    \\
29     & 159.28734 & -27.65863 & 27.039 & 0.004       & 26.488 & 0.003       & 0.551 & 1.806     & 3.04    & 1.728     & 2.95    \\
30     & 159.29143 & -27.65848 & 25.503 & 0.024       & 25.023 & 0.02        & 0.48  & 1.85      & 2.86    & 1.853     & 2.92    \\
31     & 159.2913  & -27.65848 & 26.418 & 0.079       & 25.991 & 0.055       & 0.427 & 1.967     & 3.65    & 1.849     & 3.61    \\
32     & 159.29075 & -27.65843 & 27.657 & 0.036       & 27.021 & 0.026       & 0.636 & 1.609     & 2.79    & 1.593     & 2.93    \\
33     & 159.2922  & -27.65836 & 26.634 & 0.007       & 26.027 & 0.005       & 0.607 & 1.723     & 2.77    & 1.73      & 2.92    \\
34     & 159.29242 & -27.65832 & 25.322 & 0.013       & 24.715 & 0.01        & 0.607 & 1.745     & 2.8     & 1.68      & 2.88    \\
35\footnotemark[1]     & 159.2914  & -27.6583  & 27.226 & 0.06        & 26.598 & 0.042       & 0.628 & 2.017     & 5.74    & 2.003     & 4.9     \\
36     & 159.28811 & -27.65829 & 26.071 & 0.019       & 25.617 & 0.013       & 0.454 & 1.656     & 2.87    & 1.684     & 2.96    \\
37     & 159.2915  & -27.65823 & 27.292 & 0.012       & 26.649 & 0.009       & 0.643 & 1.746     & 2.97    & 2.04      & 3.11    \\
38     & 159.29142 & -27.65821 & 26.757 & 0.017       & 26.21  & 0.012       & 0.547 & 1.774     & 2.79    & 1.858     & 2.94    \\
39     & 159.28193 & -27.6582  & 25.509 & 0.047       & 25.025 & 0.019       & 0.484 & 1.675     & 4.27    & 1.929     & 3.83    \\
40     & 159.2905  & -27.6582  & 25.874 & 0.055       & 25.212 & 0.043       & 0.662 & 1.476     & 2.78    & 1.492     & 2.89    \\
41     & 159.29438 & -27.65815 & 26.672 & 0.014       & 26.125 & 0.008       & 0.547 & 2.225     & 3.72    & 2.346     & 3.86    \\
42     & 159.28869 & -27.65812 & 26.581 & 0.038       & 25.953 & 0.025       & 0.628 & 1.516     & 2.84    & 1.571     & 2.86    \\
43     & 159.29263 & -27.6581  & 27.207 & 0.012       & 26.572 & 0.008       & 0.635 & 1.525     & 2.86    & 1.637     & 2.88    \\
44     & 159.28965 & -27.65807 & 26.017 & 0.022       & 25.518 & 0.016       & 0.499 & 1.602     & 2.85    & 1.733     & 3.17    \\
45     & 159.28919 & -27.65801 & 25.964 & 0.051       & 25.345 & 0.041       & 0.619 & 1.609     & 3.08    & 1.532     & 2.93    \\
46\footnotemark[1]      & 159.28808 & -27.658   & 27.163 & 0.041       & 26.67  & 0.029       & 0.493 & 2.037     & 3.84    & 2.389     & 3.58    \\
47     & 159.29098 & -27.65796 & 27.528 & 0.011       & 27.029 & 0.008       & 0.499 & 1.906     & 2.97    & 1.956     & 2.97    \\
48     & 159.29256 & -27.65791 & 24.911 & 0.014       & 24.323 & 0.009       & 0.588 & 1.906     & 3.84    & 1.852     & 3.33    \\
49     & 159.29104 & -27.65789 & 26.197 & 0.017       & 25.577 & 0.014       & 0.62  & 1.775     & 2.85    & 1.929     & 2.96    \\
50     & 159.28788 & -27.65784 & 25.455 & 0.009       & 24.996 & 0.006       & 0.459 & 1.633     & 2.66    & 1.738     & 2.96    \\
51     & 159.28901 & -27.65783 & 25.881 & 0.003       & 25.382 & 0.002       & 0.499 & 1.914     & 3.15    & 1.957     & 3.23    \\
52     & 159.28814 & -27.65778 & 23.951 & 0.016       & 23.481 & 0.01        & 0.47  & 1.715     & 2.77    & 1.721     & 2.89    \\
53     & 159.28799 & -27.65773 & 27.925 & 0.025       & 27.61  & 0.016       & 0.315 & 1.495     & 2.67    & 1.651     & 2.87    \\
54     & 159.28792 & -27.65765 & 26.412 & 0.041       & 25.767 & 0.033       & 0.645 & 1.409     & 2.73    & 1.421     & 2.72    \\
55     & 159.28923 & -27.65756 & 26.968 & 0.045       & 26.344 & 0.034       & 0.624 & 2.044     & 3.48    & 2.09      & 3.93    \\
56     & 159.28932 & -27.65756 & 27.535 & 0.025       & 27.131 & 0.02        & 0.404 & 2.004     & 3.06    & 1.637     & 2.99    \\
57     & 159.28942 & -27.65749 & 26.813 & 0.004       & 26.244 & 0.003       & 0.569 & 1.755     & 2.86    & 1.737     & 2.99    \\
58     & 159.29139 & -27.65747 & 26.091 & 0.039       & 25.555 & 0.032       & 0.536 & 1.375     & 2.76    & 1.294     & 2.67    \\
59     & 159.29152 & -27.65744 & 24.057 & 0.018       & 23.601 & 0.013       & 0.456 & 1.727     & 2.76    & 1.653     & 2.86    \\
60     & 159.29113 & -27.65738 & 26.964 & 0.01        & 26.537 & 0.009       & 0.427 & 1.825     & 3.11    & 1.767     & 3.16    \\
61     & 159.28876 & -27.65738 & 26.216 & 0.014       & 25.688 & 0.01        & 0.528 & 1.633     & 2.74    & 1.58      & 2.81    \\
62     & 159.29132 & -27.6573  & 25.449 & 0.055       & 25.075 & 0.04        & 0.374 & 1.522     & 2.8     & 1.45      & 2.71    \\
63     & 159.2879  & -27.65729 & 25.895 & 0.044       & 25.333 & 0.028       & 0.562 & 1.444     & 2.71    & 1.464     & 2.7     \\
64     & 159.28911 & -27.65726 & 27.452 & 0.028       & 26.886 & 0.02        & 0.566 & 1.941     & 3.05    & 2.058     & 3.43    \\
65     & 159.29074 & -27.65725 & 27.573 & 0.012       & 26.931 & 0.01        & 0.642 & 1.615     & 2.83    & 1.909     & 3.56    \\
66     & 159.29998 & -27.65715 & 26.481 & 0.004       & 25.86  & 0.002       & 0.621 & 1.684     & 2.76    & 1.718     & 2.83    \\
67     & 159.29036 & -27.65712 & 25.575 & 0.027       & 25.052 & 0.019       & 0.523 & 1.373     & 2.58    & 1.451     & 2.66    \\
68     & 159.2899  & -27.65706 & 25.463 & 0.032       & 25.003 & 0.027       & 0.46  & 1.439     & 2.6     & 1.484     & 2.7     \\
69     & 159.29694 & -27.65697 & 26.454 & 0.022       & 25.833 & 0.011       & 0.621 & 1.739     & 2.91    & 1.773     & 2.94    \\
70     & 159.28768 & -27.65692 & 26.653 & 0.007       & 26.236 & 0.005       & 0.417 & 1.871     & 3       & 1.72      & 2.84    \\
71     & 159.29066 & -27.6568  & 27.308 & 0.038       & 26.718 & 0.025       & 0.59  & 1.522     & 2.84    & 1.587     & 2.95    \\
72     & 159.29024 & -27.65671 & 25.877 & 0.032       & 25.313 & 0.025       & 0.564 & 2.193     & 4.96    & 2.043     & 5.07    \\
73     & 159.29169 & -27.65669 & 27.239 & 0.021       & 26.604 & 0.015       & 0.635 & 1.706     & 2.97    & 1.691     & 2.91    \\
74     & 159.28918 & -27.65666 & 27.037 & 0.029       & 26.573 & 0.025       & 0.464 & 1.669     & 3.01    & 1.543     & 2.88    \\
75     & 159.28757 & -27.65636 & 26.828 & 0.002       & 26.386 & 0.002       & 0.442 & 1.653     & 2.87    & 1.734     & 2.81    \\
76     & 159.28955 & -27.65627 & 26.952 & 0.053       & 26.599 & 0.029       & 0.353 & 1.636     & 3.22    & 1.888     & 3.28    \\
77     & 159.29346 & -27.65625 & 24.853 & 0.076       & 24.42  & 0.047       & 0.433 & 1.837     & 3.61    & 1.83      & 3.53    \\
78     & 159.28982 & -27.65617 & 28.507 & 0.04        & 28.028 & 0.025       & 0.479 & 1.828     & 2.95    & 2.319     & 3.58    \\
79     & 159.29111 & -27.65616 & 27.586 & 0.01        & 26.961 & 0.007       & 0.625 & 1.67      & 2.85    & 1.812     & 3.12    \\
80     & 159.29182 & -27.65609 & 26.127 & 0.004       & 25.719 & 0.003       & 0.408 & 1.75      & 2.87    & 1.862     & 3.06    \\
81     & 159.28261 & -27.64962 & 25.307 & 0.001       & 24.864 & 0.001       & 0.443 & 1.813     & 2.87    & 1.902     & 2.93    \\
82     & 159.28092 & -27.6502  & 24.554 & 0.009       & 24.151 & 0.005       & 0.403 & 1.623     & 2.95    & 1.888     & 3.15    \\
83     & 159.29015 & -27.6545  & 26.695 & 0.003       & 26.241 & 0.002       & 0.454 & 2.004     & 3.12    & 2.046     & 3.15    \\
84     & 159.29275 & -27.65174 & 25.165 & 0.018       & 24.716 & 0.01        & 0.449 & 1.678     & 2.76    & 1.858     & 3.15    \\
85     & 159.2868  & -27.65346 & 27.105 & 0.003       & 26.646 & 0.002       & 0.459 & 1.738     & 2.81    & 1.752     & 2.84    \\
86     & 159.28333 & -27.65349 & 25.44  & 0.002       & 25.001 & 0.001       & 0.439 & 1.746     & 2.78    & 1.741     & 2.81    \\
87     & 159.28723 & -27.65379 & 24.799 & 0.007       & 24.399 & 0.004       & 0.4   & 1.615     & 2.76    & 1.709     & 2.82    \\
88     & 159.28677 & -27.65439 & 26.412 & 0.022       & 25.934 & 0.012       & 0.478 & 2.139     & 3.79    & 1.753     & 2.66    \\
89     & 159.29047 & -27.65437 & 27.603 & 0.068       & 27     & 0.032       & 0.603 & 2.065     & 3.54    & 1.841     & 3.82    \\
90     & 159.29032 & -27.65482 & 24.244 & 0.001       & 23.774 & 0.001       & 0.47  & 1.696     & 2.82    & 1.724     & 2.93    \\
91     & 159.2888  & -27.65454 & 25.853 & 0.006       & 25.371 & 0.003       & 0.482 & 1.587     & 2.73    & 1.689     & 2.89    \\
92     & 159.29654 & -27.6513  & 26.89  & 0.015       & 26.398 & 0.008       & 0.492 & 1.754     & 2.91    & 1.713     & 2.98    \\
93     & 159.2899  & -27.65487 & 25.801 & 0.006       & 25.366 & 0.004       & 0.435 & 1.911     & 3.1     & 1.913     & 3.16    \\
94     & 159.29005 & -27.65557 & 25.891 & 0.007       & 25.367 & 0.004       & 0.524 & 1.725     & 2.85    & 1.761     & 2.94    \\
95     & 159.28814 & -27.6556  & 25.851 & 0.006       & 25.248 & 0.004       & 0.603 & 1.639     & 2.87    & 1.814     & 3.02    \\
96     & 159.28987 & -27.6557  & 27.162 & 0.024       & 26.538 & 0.014       & 0.624 & 1.743     & 2.72    & 1.554     & 2.86    \\
97     & 159.29037 & -27.65583 & 26.216 & 0.01        & 25.62  & 0.006       & 0.596 & 1.665     & 2.86    & 1.683     & 2.98    \\
98     & 159.29296 & -27.65223 & 24.58  & 0.002       & 24.181 & 0.001       & 0.399 & 1.88      & 2.91    & 1.924     & 3.06    \\
99     & 159.28485 & -27.6556  & 24.649 & 0.001       & 24.119 & 0.001       & 0.53  & 1.559     & 2.63    & 1.588     & 2.56    \\
100    & 159.29129 & -27.65171 & 26.722 & 0.013       & 26.263 & 0.007       & 0.459 & 1.847     & 3.12    & 2.027     & 3.66    \\
101    & 159.29226 & -27.652   & 26.458 & 0.01        & 26.041 & 0.006       & 0.417 & 1.651     & 2.65    & 1.713     & 2.89    \\
102    & 159.28552 & -27.65198 & 26.045 & 0.005       & 25.638 & 0.003       & 0.407 & 1.954     & 3.37    & 1.865     & 2.94 \\

\footnotetext[1]{These objects (\#35 and \#46) are the "faint fuzzies" candidates.}
\end{longtable}
\end{landscape}

\end{appendix}
\end{document}